\documentclass[sigchi]{acmart}
\usepackage{tabularx} % For formal tables
\usepackage{booktabs} % For formal tables
\usepackage[ruled]{algorithm2e} % For algorithms
\usepackage[colorinlistoftodos]{todonotes}
\usepackage{verbatim}
\usepackage{graphicx}
\usepackage{subfig}

\usepackage{listings}

\usepackage{multirow}
\usepackage[normalem]{ulem}

\SetAlFnt{\small}
\SetAlCapFnt{\small}
\SetAlCapNameFnt{\small}
\SetAlCapHSkip{0pt}
\IncMargin{-\parindent}

\def\statisticalAnalysis{statistical analysis\xspace}

\def\statisticalAnalyses{statistical analyses\xspace}

\def\TeaPL{Tea's programming language\xspace}
\def\TeaRS{Tea's runtime system\xspace}

\def\dataSet{data set\xspace}

\newcommand\colH[1]{\multicolumn{1}{c}{\textbf{#1}}}

\newcolumntype{H}{>{\setbox0=\hbox\bgroup}c<{\egroup}@{}}

\newcommand{\assume}[1]{\tikz[baseline]\node[circle,inner xsep=2pt,inner ysep=0pt,
        draw=black,fill=white,scale=0.8,anchor=base] {#1};}
\newcommand{\valid}[1]{\tikz[baseline]\node[circle,inner xsep=2pt,inner ysep=0pt,
        draw=black,fill=blue!30!white,scale=0.8,anchor=base] {#1};}
\newcommand{\invalid}[1]{\tikz[baseline]\node[circle,inner xsep=2pt,inner ysep=0pt,
        draw=black,fill=red!30!white,scale=0.8,anchor=base] {#1};}
\def\yes{$\checkmark$}
\def\no{---}

\def\<#1>{\codeid{#1}}
\newcommand{\codeid}[1]{\ifmmode{\mbox{\small\ttfamily{#1}}}\else{\small\ttfamily #1}\fi}
\newcommand{\codeidsmall}[1]{\ifmmode{\mbox{\smaller\ttfamily{#1}}}\else{\smaller\ttfamily #1}\fi}

\acmConference[arxiv]{\url{arxiv.org}}{April 2019}{on-line}
\acmYear{2019}
\setcopyright{iw3c2w3}
\acmDOI{0000001.0000001}

\usepackage{cuted}
\usepackage{capt-of}

\hypersetup{draft}

\begin{document}
% Title portion
\title{Tea: A High-level Language and Runtime System \\ for Automating Statistical Analyses}

\author{Eunice Jun, Maureen Daum, Jared Roesch}
% \orcid{1234-5678-9012-3456}
\affiliation{%
  \institution{University of Washington}
  \streetaddress{}
  \city{}
  \state{}
  \postcode{}
  \country{}}
\email{}

\author{Sarah E. Chasins}
% \orcid{1234-5678-9012-3456}
\affiliation{%
  \institution{University of California, Berkeley}
  \streetaddress{}
  \city{}
  \state{}
  \postcode{}
  \country{}}
\email{}

\author{Emery D. Berger}
% \orcid{1234-5678-9012-3456}
\affiliation{%
  \institution{University of Massachusetts Amherst}
  \streetaddress{}
  \city{}
  \state{}
  \postcode{}
  \country{}}
\email{}

\author{Rene Just, Katharina Reinecke}
% \orcid{1234-5678-9012-3456}
\affiliation{%
  \institution{University of Washington}
  \streetaddress{}
  \city{}
  \state{}
  \postcode{}
  \country{}}
\email{}

\newcommand{\figureDesignConsiderations}{
\begin{figure}[t]
	\vspace{-5pt}
    \centering
	\includegraphics[width=1.\columnwidth, clip]{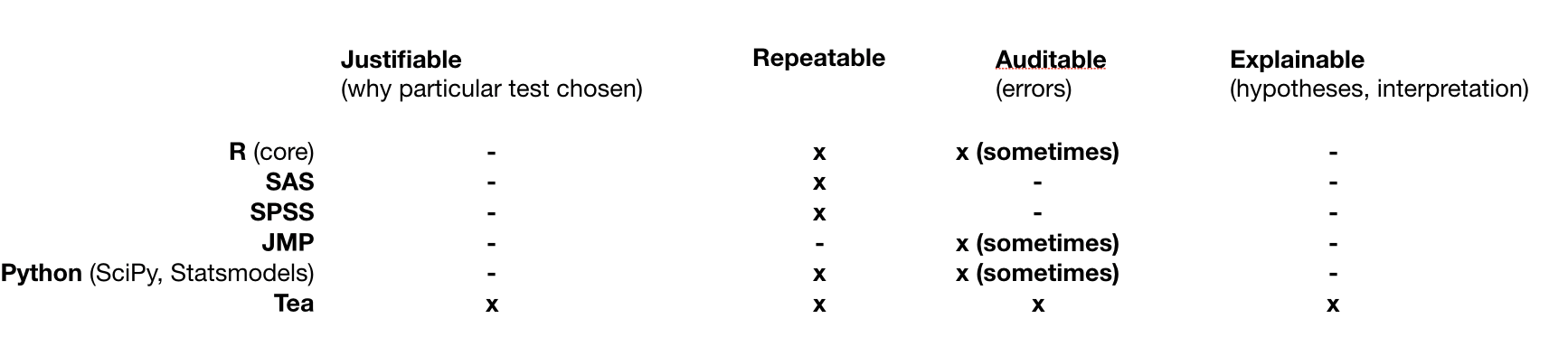}
	\vspace{-7pt}
    \caption{Mockup of design considerations}
    \label{fig:designConsiderations}
    \vspace{-15pt}
\end{figure}
}

\newcommand{\figureCEGIS}{
\begin{figure}[t]
	% \vspace{-5pt}
    \centering
	\includegraphics[width=1.\columnwidth, clip]{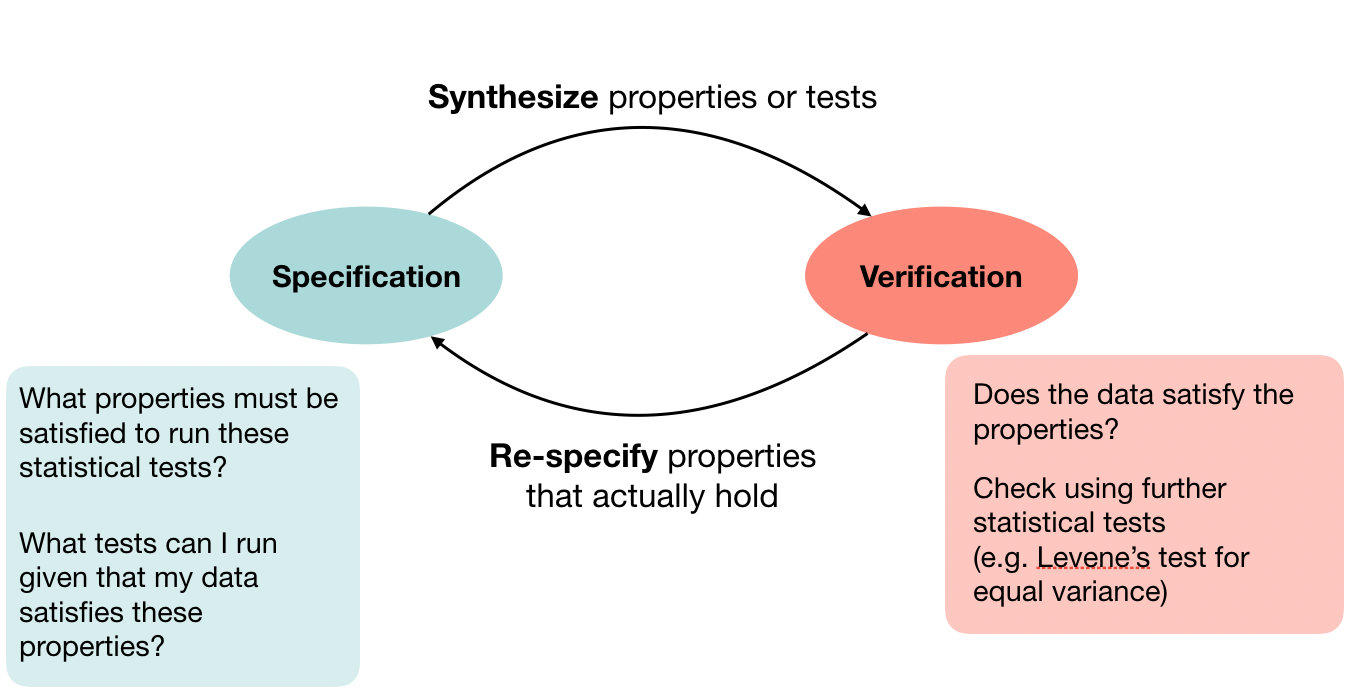}
	\vspace{-7pt}
    \caption{Tea's counterexample-guided inductive synthesis approach.}
    % synthesize a set of valid statistical tests based on user assumptions. If users'
    % assumptions are partial (only address a subset of properties pertaining to a
    % test), Tea verifies the other properties to determine if a test really does
    % hold. If users specify desired statistical tests, then Rooibos synthesizes the
    % seof properties that must hold in the data in order to run the tests. If thtests
    % have contradictory properties (e.g., variables cannot be botcontinuous and
    % categorical), Rooibos surfaces this contradiction anprovides classes of tests
    % and properties that are compatible with eacother.}
    \label{fig:cegis}
    \vspace{-15pt}
\end{figure}
}

\newcommand{\figureTestsToProperties}{
\begin{figure}[t]
    \vspace{-5pt}
    \centering
    \includegraphics[width=1.\columnwidth, clip]{figures/tests_to_properties.png}
    \vspace{-7pt}
    \caption{Output of a Tea program that computes the set of properties necessary to run a set of tests. In this case, the tests that are specified require complementary properties, so all constraints can be satisfied.}
\end{figure}
}

\newcommand{\figureTeaProgram}{
\begin{figure}[t]
    \vspace{-5pt}
    \centering
    \includegraphics[width=1.\columnwidth, clip]{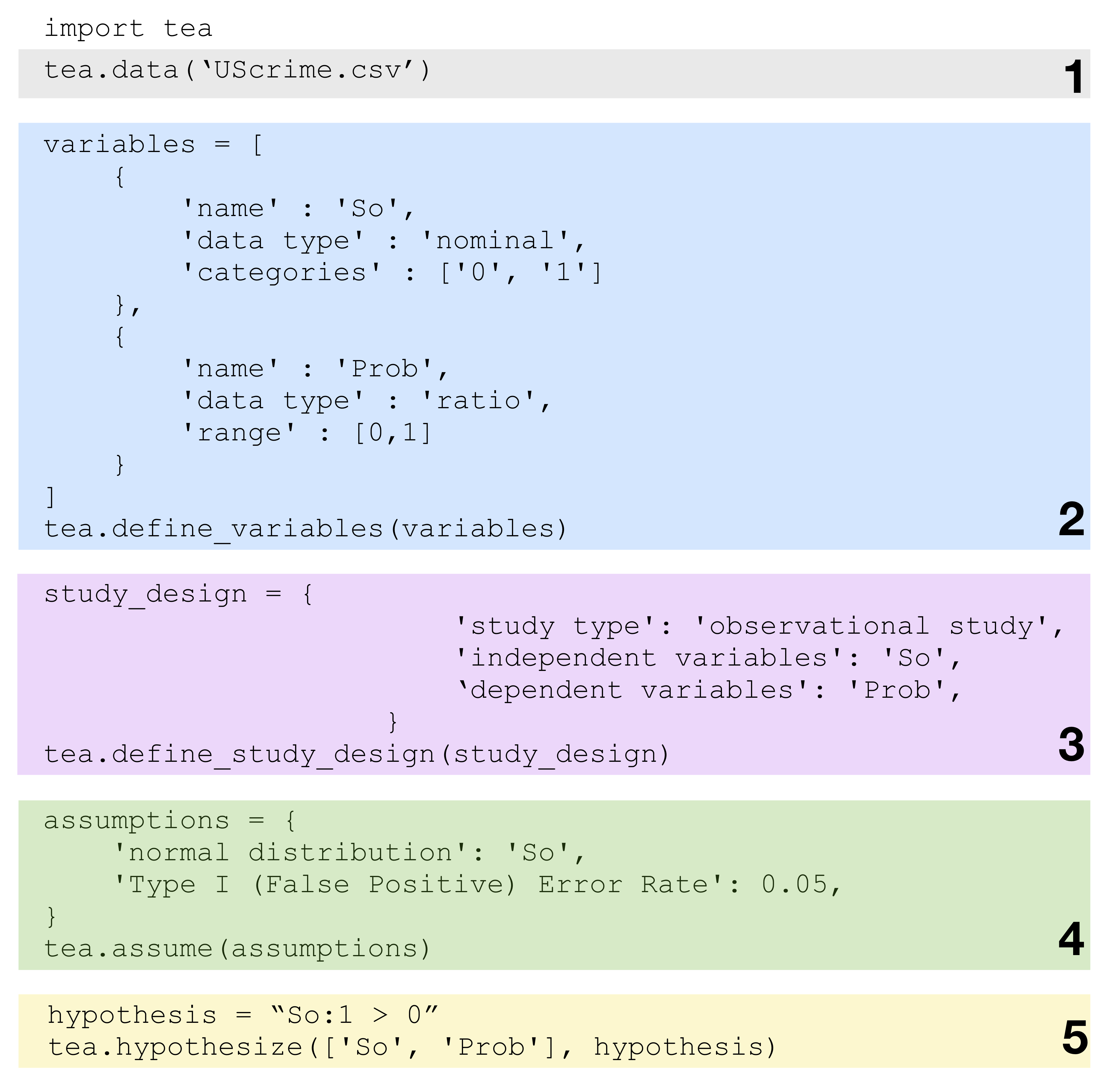}
    \vspace{-7pt}
    \caption{Sample Tea program specification that outlines an experiment to analyze the relationship between geographic location (`So') and probability of imprisonment (`Prob') in a common USCrime dataset~\cite{venables2013modern, kabacoff2011action}. See Section~\ref{sec:usagescenario} for an explanation of the code. 
    Tea programs specify 1) data, 2) variables, 3) study design, 4) assumptions, and 5) hypotheses.}
    \label{fig:tea_program}
\end{figure}
}

\newcommand{\figureTeaser}{
    \begin{strip}
        \centering
        \includegraphics[width=\textwidth]{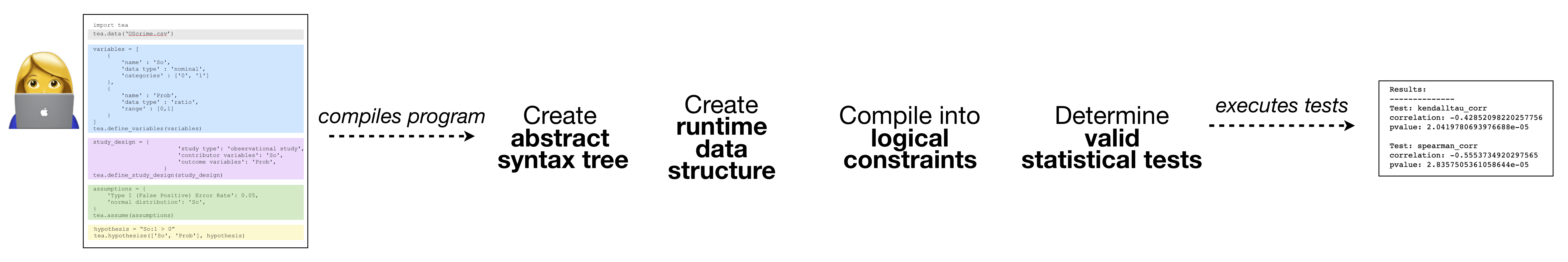}
        \captionof{figure}{Overview of Tea's language constructs and its runtime system's compilation process. The colors correspond to the annotations provided in Figure~\ref{fig:tea_program}.  
        % \chasins{If there's time, make verbs match in fig (`compiles' -> `compile', `executes' -> `execute').}
        \label{fig:overview}}
    \end{strip}
}

% \newcommand{\figureTeaOverview}{
%     \begin{figure}[!tbp]
%         \centering
%         \subfloat[Sample Tea program.]{\includegraphics[width=0.4\textwidth]{figures/tea_program.png}\label{fig:program_small}}
%         \hfill
%         \subfloat[Tea and RooiboS.]{\includegraphics[width=0.4\textwidth]{figures/tea_overview.png}\label{fig:overview}}
%         \caption{Overview of Tea's API and execution process}
%         \label{fig:tea_overview}
% \end{figure}
% }

% \newcommand{\figureTeaOverview}{
%     \begin{figure}[!tbp]
%         \centering
%         \includegraphics[width=0.4\textwidth]{figures/tea_overview.png}\label{fig:overview}}
%         \caption{Overview of Tea's API and execution process}
%         \label{fig:tea_overview}
% \end{figure}
% }

\def\r{Pearson's r}
\def\ktau{Kendall's $\tau$}
\def\srho{Spearman's $\rho$}
\def\pb{Pointbiserial}
\def\student{Student's t-test}
\def\paired{Paired t-test}
\def\mannu{Mann-Whitney U}
\def\wilcox{Wilcoxon signed rank}
\def\welch{Welch's}
\def\f{F-test}
\def\rm{Repeated measures one way ANOVA}
\def\kw{Kruskal Wallis}
\def\friedman{Friedman}
\def\facANOVA{Factorial ANOVA}
\def\twoANOVA{Two-way ANOVA}
\def\chiSq{Chi Square}
\def\fisher{Fisher's Exact}

\newcommand{\tableTeaTests}{
    \begin{table*}[htbp]
      \begin{center}
      \caption{Statistical tests supported in Tea's Null Hypothesis Significance Testing module} \label{tab:tea_tests}
      \begin{tabular}{lll}
      \toprule
      \colH{Class of tests}           & \colH{Parametric} & \colH{Non-parametric} \\
        Correlation                   & \r                &  \ktau   \\
                                      & \pb               &  \srho  \\
        \midrule
        Bivariate mean comparison     & \student          & \welch \\
                                      &                   & \mannu \\
                                      &                   & (a.k.a. Wilcoxon rank sum) \\
                                      & \paired           & \wilcox \\ 
        \midrule
        Multivariate mean comparison  & \f                & \kw   \\
                                      & \rm               & \friedman \\
                                      & \twoANOVA         & \\
                                      & \facANOVA         & \\
        \midrule
        \midrule
        Proportions: \chiSq , \fisher \\
        Others: Bootstrapping (with confidence intervals) \\
      \bottomrule
      \end{tabular}
      \end{center}
    \end{table*}
}

\newcommand{\furtherTestResults}{
\begin{table*}[htbp]
  \begin{center}
    \caption{Results from Factorial ANOVA for RM ANOVA example.}
    \begin{tabularx}{\linewidth}{XXXXXX}
       & \textbf{df} & \textbf{sum\_sq} & \textbf{mean\_sq} & \textbf{F} & \textbf{PR(>F)} \\
      C(conc) & 6.0 & 4068.771429 & 678.128571 & 9.261087 & 1.242777e-07 \\
      Residual & 77.0 & 5638.204167  & 73.223431   &    NaN     &      NaN \\
    \end{tabularx}
    \end{center}
    \label{tab:rm_anova_fanova}
\end{table*}

\begin{table*}[htpb]
  \begin{center}
    \caption{Results from F test for F test example.}
    \begin{tabularx}{\linewidth}{XXXXXX}
       & \textbf{df} & \textbf{sum\_sq} & \textbf{mean\_sq} & \textbf{F} & \textbf{PR(>F)} \\
        C(trt)   &  4.0 & 1351.369014 & 337.842253 & 32.432826 & 9.818516e-13 \\
        Residual & 45.0 &  468.750438  & 10.416676    &    NaN       &    NaN \\
    \end{tabularx}
  \end{center}
  \label{tab:f_test_f_test}
\end{table*}

\begin{table*}[htpb]
  \begin{center}
    \caption{Results from Factorial ANOVA test for F test example.}
    \begin{tabularx}{\linewidth}{XXXXXX}
       & \textbf{df} & \textbf{sum\_sq} & \textbf{mean\_sq} & \textbf{F} & \textbf{PR(>F)} \\
        C(trt)   &  4.0 & 1351.369014 & 337.842253 & 32.432826 & 9.818516e-13 \\
        Residual & 45.0 &  468.750438  & 10.416676    &    NaN       &    NaN \\
    \end{tabularx}
  \end{center}
  \label{tab:f_test_fanova}
\end{table*}

\begin{table*}[htpb]
  \begin{center}
    \caption{Results from Factorial ANOVA test for Paired T test example.}
    \begin{tabularx}{\linewidth}{XXXXXX}
       & \textbf{df} & \textbf{sum\_sq} & \textbf{mean\_sq} & \textbf{F} & \textbf{PR(>F)} \\
        C(Group) &  1.0 &  294.0  &  294.0 & 2.826923 & 0.106839 \\
        Residual & 22.0 & 2288.0  &  104.0    &   NaN   &    NaN \\
    \end{tabularx}
  \end{center}
  \label{tab:paired_t_fanova}
\end{table*}

\begin{table*}[htpb]
  \begin{center}
    \caption{Results from F test and Factorial ANOVA for Student's T test example.}
    \begin{tabularx}{\linewidth}{XXXXXX}
       & \textbf{df} & \textbf{sum\_sq} & \textbf{mean\_sq} & \textbf{F} & \textbf{PR(>F)} \\
        C(So)   &   1.0 &  0.006702 &  0.006702 & 17.657903 & 0.000124 \\
        Residual & 45.0 & 0.017079 & 0.000380    &    NaN     &  NaN \\
    \end{tabularx}
  \end{center}
  \label{tab:students_t_f_test_and_fanova}
\end{table*}
}

% make a custom style that looks good and can highlight some additional keywords
\lstdefinestyle{tea}{
  basicstyle=\ttfamily,
  deletekeywords={input, print, id},
  language=Python,
  % here are the additional keywords
  emph={load_data, hypothesize, <, >, =, !},
  % they are underlines
  emphstyle={\bf},
}
\lstset{
  % use this style by default
  style=tea,
  % look better
  columns=flexible,
  showstringspaces=false,
  % spacing, size, numbers, etc.
  numbers=left,
  xleftmargin=2em,
  numberstyle=\tiny,
  escapechar=|,
}
\newcommand{\listHypotheses}{
    \lstinputlisting[caption=Hypothesis grammar]{figures/hypotheses.py}
    \label{list:hypotheses}
}

\newcommand{\teaHypotheses}{
  \begin{figure}
    \vspace{-5pt}
    \centering
    \includegraphics[width=1.\columnwidth, clip]{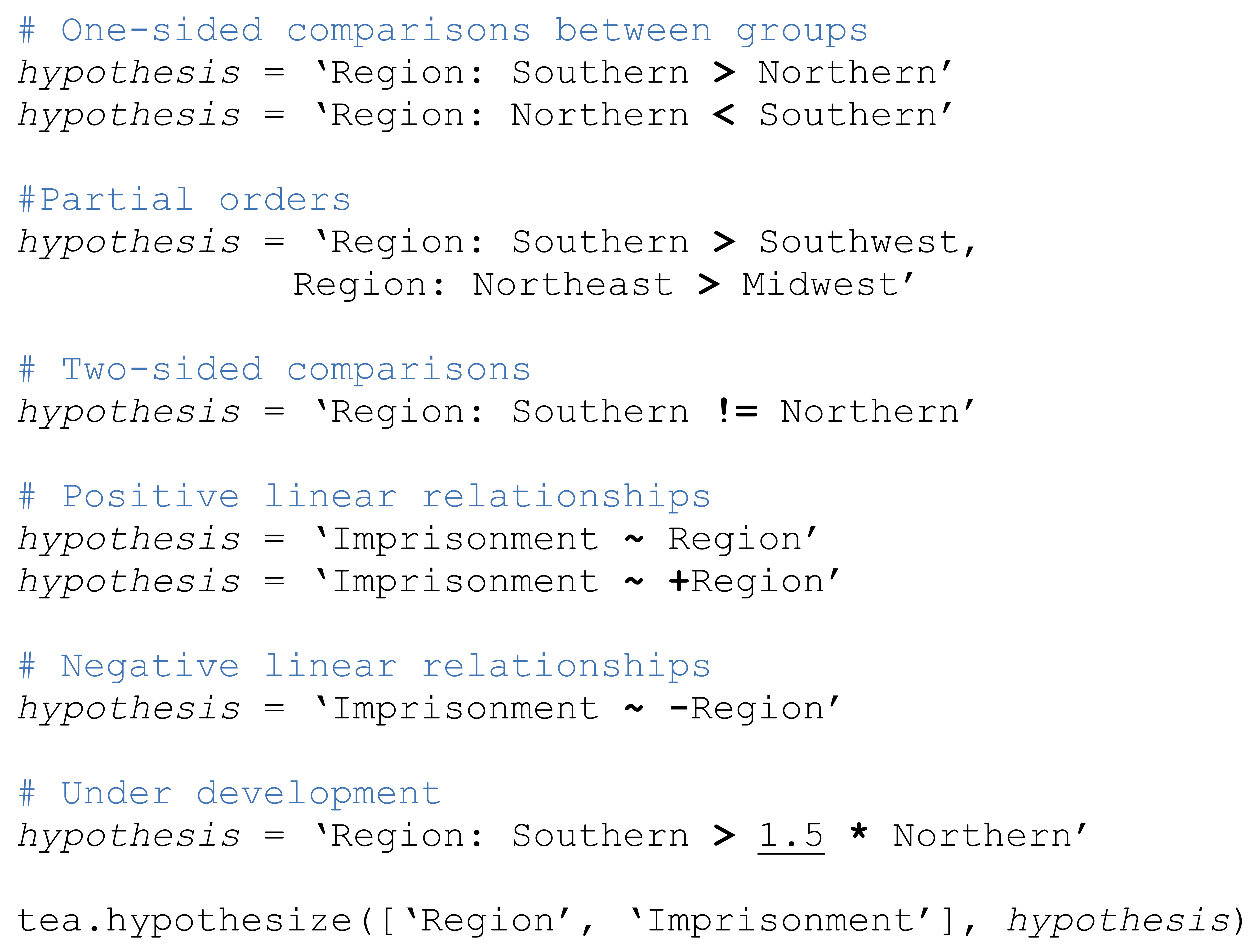}
    \vspace{-7pt}
    \caption{Hypotheses that users can express in Tea. }
    \label{fig:teaHypotheses}
    \vspace{-15pt}
  \end{figure}
}

\newcommand{\figureRTeaComparison}{
\begin{figure}
    \vspace{-5pt}
    \centering
    \includegraphics[width=1.\columnwidth, clip]{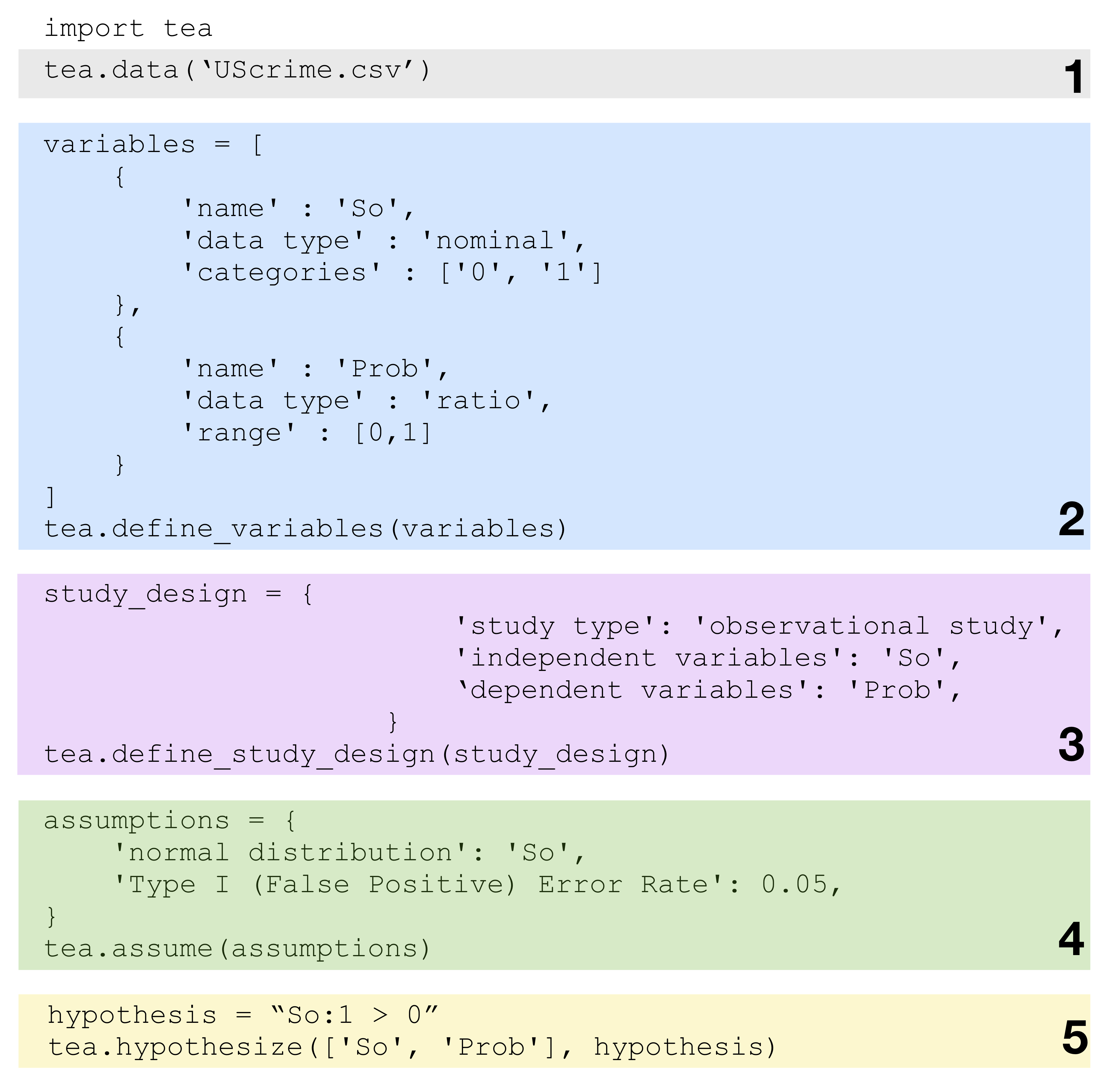}
    \vspace{-7pt}
    \caption{Comparison of R and Tea programs. The figure needs to be udpated.}
    \label{fig:rTeaComparison}
\end{figure}
}

\newcommand{\otherSystems}{
    \begin{table*}[t]
        \centering
        \caption{Comparison of Tea to other tools. Despite the published best practices for statistical analyses, most tools do not help users select appropriate tests. Tea not only addresses the best practices but also supports reproducing analyses.}
        \label{tab:otherSystems}
        \begin{tabularx}{\linewidth}{r|c|c|c|c|c|c}
            \colH{Best practices} & \colH{SAS} & \colH{SPSS} & \colH{JMP} & \colH{R} & \colH{Statsplorer} & \colH{Tea} \\
            Explicit statement of user assumptions & \no & \no & \no & \no & \no & \yes \\
            Automatic verification of test preconditions & \no & \no & sometimes & sometimes & \yes & \yes \\
            Automatic accounting of multiple comparisons & \no & \no & \no & \no & \yes & \yes \\ 
            Surface alternative analyses & \no & \no & \no & \no & \no & \yes \\
            Contextualize results & \yes & sometimes & \yes & sometimes & \yes & \yes \\
            Easy to reproduce analysis & \yes & \yes & \no & \yes & \no & \yes \\
        \end{tabularx}
    \end{table*}
}

\figureTeaser
\begin{abstract}
Though \statisticalAnalyses are centered on research questions and
hypotheses, current statistical analysis tools are not. Users must
first translate their hypotheses into specific statistical tests and
then perform API calls with functions and parameters. To do so
accurately requires that users have statistical expertise. To lower
this barrier to valid, replicable \statisticalAnalysis, we introduce
Tea\footnote{named after Fisher's ``Lady Tasting Tea'' experiment~\cite{fisher1937design}}, a high-level declarative language and runtime system.  In Tea,
users express their study design, any parametric assumptions, and
their hypotheses. Tea compiles these high-level specifications into a
constraint satisfaction problem that determines the set of valid
statistical tests, and then executes them to test the hypothesis. We
evaluate Tea using a suite of statistical analyses drawn from popular
tutorials. We show that Tea generally matches the choices of experts
while automatically switching to non-parametric tests when parametric
assumptions are not met. We simulate the effect of mistakes made by
non-expert users and show that Tea automatically avoids both false
negatives and false positives that could be produced by the application of
incorrect statistical tests.

\end{abstract}




\maketitle

\section{Introduction} \label{sec:Introduction}
% Problem: People know their questions and hypotheses but have a hard time
% selecting the correct statistical tests to test their questions and hypotheses. 

% We provide formal semantics to statistical analysis. With this semantics, we are
% able to provide a high-level declarative language users can express dmoain knowledge, 
\begin{quote}
\textit{
The enormous variety of modern quantitative methods leaves researchers with the
nontrivial task of matching analysis and design to the research question.
                                - Ronald Fisher~\cite{fisher1937design}}
\end{quote}

Since the development of modern statistical methods (e.g., Student's
t-test, ANOVA, etc.), statisticians have acknowledged the difficulty
of identifying which statistical tests people should use to answer
their specific research questions.
% \chasins{individual methods to individual questions?  whether some methods should 
% be applied to research questions at all?  open with clear statement of what tea's tackling}.  
Almost a century later, choosing appropriate statistical tests for
evaluating a hypothesis remains a challenge. As a consequence, errors
in statistical analyses are common~\cite{kaptein2012rethinking},
especially given that data analysis has become a common task for
people with little to no statistical expertise.

%Data analysis has become a common task for people with little to nostatistical expertise. 
A wide variety of tools (such as SPSS~\cite{wiki:spss},
SAS~\cite{wiki:sas}, and JMP~\cite{wiki:jmp}), programming languages
(R~\cite{wiki:r-language}), and libraries (including
numpy~\cite{oliphant2006numpy}, scipy~\cite{scipy}, and
statsmodels~\cite{seabold2010statsmodels}), enable people to perform
specific statistical tests, but they do not address the fundamental
problem that users may not know which statistical test to perform and
how to verify that specific assumptions about their data hold.
% \kr{Alternative start of
% the intro: In ``The Lady Tasting Tea'', the author David Salsburg vividly
% describes the evolution of statistics, from Fisher's theories to .... . But
% besides introducing the foundations of statistics, the book illustrates the
% uncertainty in the appropriate use of statistics -- something that still remains
% an issue today. 
% In fact, users of statistical tests still face the burden of deciding which test to use, whether it satisfies a set of assumptions about the data, and how to interpret the results.... }

In fact, all of these tools place the burden of valid, replicable
statistical analyses on the user and demand deep knowledge of
statistics.
%Currenttools require users to be aware of statistical details. 
Users not only have to identify their research questions, hypotheses,
and domain assumptions, but also must select statistical tests for
their hypotheses (e.g., Student's t-test or one-way ANOVA). For each
statistical test, users must be aware of the statistical assumptions
each test makes about the data (e.g., normality or equal variance
between groups) and how to check for them, which requires additional
statistical tests (e.g., Levene's test for equal variance), which
themselves may demand further assumptions about the data. This entire
process requires significant knowledge about statistical tests and
their preconditions, as well as the ability to perform the tests and
verify their preconditions. This cognitively demanding process can
easily lead to mistakes.
% and the correct interpretation of  results. 
%Throughout this process, the possibility that users overlook  equally valid alternative tests is present and unavoidable. Being aware of the analysis goals and the details of statistical tests and their preconditions is challenging. 

This paper presents Tea, a high-level declarative language for
automating statistical test selection and execution that abstracts the
details of statistical analysis from the users. Tea captures users'
hypotheses and domain knowledge, translates this information into a
constraint satisfaction problem, identifies all valid statistical
tests to evaluate a hypothesis, and executes the tests.
Figure~\ref{fig:overview} illustrates Tea's compilation process. Tea's
higher-level, declarative nature aims to lower the barrier to valid,
replicable analyses.

We have designed Tea to integrate directly into common data analysis
workflows for users who have minimal programming experience. Tea is
implemented as an open-source Python library, so programmers can use
Tea wherever they use Python, including within Python notebooks.

In addition, Tea is flexible. Its abstraction of the analysis process
and use of a constraint solver to select tests is designed to support
its extension to emerging statistical methods, such as Bayesian
analysis. Currently, Tea supports frequentist Null Hypothesis
Significance Testing (NHST).

The paper makes the following contributions:
\begin{itemize}
    \item Tea, a novel domain-specific language (DSL) for automatically selecting and
    executing statistical analyses based on users' hypotheses and domain
    knowledge (~\autoref{sec:TeaPL}), 
    \item the Tea runtime system, which formulates statistical test selection as a maximum constraint satisfaction problem (~\autoref{sec:TeaRS}), and
    \item an initial evaluation showing that Tea can express and execute common NHST statistical tests (~\autoref{sec:eval}). 
\end{itemize}

We start with a usage scenario that provides an overview of Tea
(\autoref{sec:usagescenario}). We discuss the concerns about
statistics in the HCI community that shaped Tea's
design~(\autoref{sec:design}), the implementation of
\TeaPL~(\autoref{sec:TeaPL}), the implementation of
\TeaRS~(\autoref{sec:TeaRS}), and the evaluation of Tea as a
whole~(\autoref{sec:eval}). We then discuss limitations and future
work and how Tea is different from related work. We conclude by
providing information on how to use Tea.

\section{Usage Scenario}\label{sec:usagescenario}

This section describes how an analyst who has no statistical
background can use Tea to answer their research questions. We use as
an example analyst a historical criminologist who wants to determine
how imprisonment differed across regions of the US in
1960\footnote{The example is taken
  from~\cite{ehrlich1973participation}
  and~\cite{vandaele1987participation}. The data set comes as part of
  the MASS package in R.}. Figure~\ref{fig:tea_program} shows the Tea
code for this example.

The analyst specifies the data file's path in Tea. Tea handles loading and
storing the data set for the duration of the analysis session. The analyst does
not have to worry about transforming the data in any way. 

The analyst asks if the probability of imprisonment was higher in
southern states than in non-southern states. The analyst identifies
two variables that could help them answer this question: the
probability of imprisonment (`Prob') and geographic location
(`So'). %The latter has already been coded as `1' for southern and `0'
for non-southern.  Using Tea, the analyst defines the geographic
location as a dichotomous nominal variable where `1' indicates a
southern state and `0' indicates a non-southern state, and indicates that the
probability of imprisonment is a numeric data type (ratio) with a
range between 0 and 1. %The analyst can additionally identify any
%other variables they care about in the data set.

The analyst then specifies their study design, defining the study type
to be `observational study' (rather than experimental study) and
defining the independent variable to be the geographic location and
the outcome (dependent) variable to be the probability of
imprisonment.

Based on their prior research, the analyst knows that the probability
of imprisonment in southern and non-southern states is normally
distributed. The analyst provides an assumptions clause to Tea in
which they specify this domain knowledge. They also specify an
acceptable Type I error rate (probability of finding a false positive
result), more colloquially known as the `significance threshold'
($\alpha = .05$) that is acceptable in criminology. If the analyst
does not have assumptions or forgets to provide assumptions, Tea will
use the default of $\alpha = .05$.

% Should we switch the story where hypothesis turns out to be incorrect and alt
% metrics give evidence of that?
The analyst hypothesizes that southern states will have a higher
probability of imprisonment than non-southern states. The analyst
directly expresses this hypothesis in Tea.  \emph{Note that at no
  point does the analyst indicate which statistical tests should be
  performed.}

From this point on, Tea operates entirely automatically.  When the
analyst runs their Tea program, Tea checks properties of the data and
finds that Student's t-test is appropriate. Tea executes the Student's
t-test and non-parametric alternatives, such as the Mann-Whitney U
test, which provide alternative, consistent results.

Tea generates a table of results from executing the tests, ordered by their
power (i.e., results from the parametric t-test will be listed first
given that it has higher power than the non-parametric
equivalent). Based on this output, the analyst concludes that their
hypothesis---that the probability of imprisonment was higher in
southern states than in non-southern states in 1960---is
supported. The results from alternative statistical tests support this
conclusion, so the analyst can be confident in their assessment.

%The analyst wants to conduct the same analysis with a data set from a different
%year. In Tea, they have to change only one line of code: the file path. As long as the variables exist in the new data set, Tea can conduct the same analysis without altering the rest of the code.\chasins{is this very different from prior approaches?  say python?  does it warrant this much emphasis?}

The analyst can now share their Tea program with colleagues.  Other
researchers can easily see what assumptions the analyst made and what
the intended hypothesis was (since these are explicitly stated in the
Tea program), and reproduce the exact results using Tea.

% If the analyst wanted to do the same analysis for multiple years,
% the analyst could use an outer Python loop to execute the same analysis for all data
% sets they had. 

%Tea enables users who may not have statistical expertise to conduct valid, replicable analyses without having to write statistical functions \chasins{clarify}. Instead, users focus on expressing knowledge about their data source, variables of interest, assumptions, and hypotheses. Tea automates test selection, test precondition checking, and test execution, and it surfaces multiple valid statistical tests for testing a given hypothesis. Tea analyses can also be shared and re-run. 

%\section{Statistics in HCI} \label{sec:statsHCI}
\otherSystems
\section{Design Considerations} \label{sec:design}

%The American Psychological Assocation (APA) initiated a Task Force on Statistical Inference~\cite{APATFSI} in the late 1990s to address concerns about statistical practices~\cite{wilkinson1999statistical}. 
% The guidelines and recommendations for
% study design, analysis, and reporting

%In 2007, Cairns outlined common statistical analysis problems in the HCI community that echo concerns articulated in~\cite{wilkinson1999statistical}: not checking or reporting the assumptions made by statistical tests, choosing the incorrect statistical tests to test hypotheses, conducting statistical tests multiple times (multiple comparison), and inconsistent reporting of results, including the omission of non-statistically significant results. 

In designing Tea's language and runtime system, we considered best practices for conducting statistical analyses and derived our own insights on improving the
interaction between users and statistical tools.

We identified five key recommendations for statistical analysis from Cairns' report on common
statistical errors in HCI~\cite{cairns2007hci}, which echoes many concerns articulated by Wilkinson~\citet{wilkinson1999statistical}, and from the American Psychological Association's
Task Force on Statistical Inference~\cite{APATFSI}: 
\begin{itemize}
    \item Users should make their assumptions about the data explicit~\cite{APATFSI}. 
    \item Users should check assumptions statistical tests make about the data and variables and report on the results from conducting tests to check these assumptions~\cite{cairns2007hci,APATFSI}.
    \item Users should account for multiple comparisons~\cite{cairns2007hci,APATFSI}.
    \item When possible, users should consider alternative analyses that test their hypothesis and select the simplest one~\cite{APATFSI}.
    \item Users should contextualize results from statistical tests using effect sizes and confidence intervals~\cite{APATFSI}.
\end{itemize}

An additional practice we wanted to simplify in Tea was \textit{reproducing analyses}. Table~\ref{tab:otherSystems} shows how Tea compares to current tools in supporting these best practices.

% The last four recommendations pertain to details that require statistical
% expertise many users may not have. Tea aims to lower the barrier to valid statistical analysis. 

Based on these guidelines, we identified two key interaction principles for Tea: 
\begin{enumerate}
    \item \textit{Users should be able to express their expertise, assumptions,
    and intentions for analysis.} Users have domain knowledge and goals
    that cannot be expressed with the low-level API calls to the specific
    statistical tests required by the majority of current tools. A higher level
    of abstraction that focuses on the goals and context of analysis is
    likely to appeal to users who may not have statistical expertise (\autoref{sec:TeaPL}).
    \item \textit{Users should not be burdened with statistical details to
    conduct valid analyses.} Currently, users must not only remember their hypotheses but
    also identify possibly appropriate tests and manually check the
    preconditions for all the tests. %best practices and steps to data analysis.
    Simplifying the user's procedure by automating the test selection process
    can help reduce cognitive demand (\autoref{sec:TeaRS}).
\end{enumerate}

While there are calls to incorporate other methods of statistical
analysis~\cite{kay2016researcher,kaptein2012rethinking}, Null
Hypothesis Significance Testing (NHST) remains the norm in HCI and
other disciplines. Therefore, Tea currently implements a module for
NHST with the tests found to be most common
by~\citet{wacharamanotham2015statsplorer} (see
Table~\ref{tab:tea_tests} for a list of tests). We believe that Tea's
abstraction and modularity will enable the incorporation of other
statistical analysis approaches as they move into the mainstream.

\section{Tea's Programming Language} \label{sec:TeaPL}
\figureTeaProgram
Tea is a domain-specific language
embedded in Python. It takes advantage of existing Python data structures (e.g.,
classes, dictionaries, and enums). We chose Python because of its widespread
adoption in data science. Tea is itself implemented as a Python library.
%\footnote{We are open-sourcing Tea and
%will be making it installable using \texttt{pip}, a common Python package manager.}. 

%Users interact with Tea through a high-level, declarative API that
%internally builds an abstract syntax tree (AST). The nodes in the AST
%correspond to variables and their relationships with one another.

% We designed Tea's language with a few key concerns in mind: minimizing user
% cognitive load, reducing user's manipulation and duplication of data, and XX. 
% (Relate to design considerations....)
% (in R with all the dataframes and subsets of
% dataframes) A key concern in designing the language was in providing a simple
% interface that would minimize user cognitive load, 

A key challenge in describing studies is determining the level of
granularity necessary to produce an accurate analysis.  In Tea
programs, users describe their studies in five ways: (1) providing a
data set, (2) describing the variables of interest in that \dataSet,
(3) describing their study design, (4) explicitly stating their
assumptions about the variables, and (5) formulating hypotheses about
the relationships between variables.

\subsection{Data}
Data is required for executing statistical analyses. One challenge in managing
data for analysis is minimizing both duplicated data and user intervention.

To reduce the need for user intervention for data manipulation, Tea
requires the data to be a CSV in long format. CSVs are a common output
format for data storage and cleaning tools. Long format (sometimes
called ``tidy data''~\cite{wickham2014tidy}) is a denormalized format
that is widely used for collecting and storing data, especially for
within-subjects studies.

Unlike R and Python libraries such as numpy~\cite{oliphant2006numpy}, Tea only
requires one instance of the data. Users do not have to duplicate the data or
subsets of it for analyses that require the data to be in slightly different
forms. Minimizing data duplication or segmentation is also important to avoid
user confusion about where some data exist or which subsets of data pertain to
specific statistical tests.

Optionally, users can also indicate a column in the \dataSet that acts
as a relational (or primary) key, or an attribute that uniquely
identifies rows of data. For example, this key could be a participant
identification number in a behavioral experiment. A key is useful for
verifying a study design, described below. Without a key, Tea's default
is that all rows in the dataset comprise independent observations (that is, all
variables are between subjects).

\subsection{Variables}
Variables represent columns of interest in the data set. Variables
have a name, a data type (\emph{nominal}, \emph{ordinal},
\emph{interval}, or \emph{ratio}), and, when appropriate, valid
categories.  Users (naturally) refer to variables through a Tea program using
their names. Only nominal and ordinal variables have a list of
possible categories. For ordinal variables, the categories are also
ordered from left to right.

Variables encapsulate queries. The queries represent the index of the
variable's column in the original data set and any filtering
operations applied to the variable. For instance, it is common to
filter by category for nominal variables in statistical tests.

\subsection{Study Design}
Three aspects of study design are important for conducting statistical
analyses: (1) the type of study (observational study vs. randomized
experiment), (2) the independent and dependent variables, and (3) the
number of observations per participant (e.g., between-subjects
variables vs. within-subjects variables).

For semantic precision, Tea uses different terms for independent and
dependent variables for observational studies and experiments.  In
experiments, variables are described as either ``independent'' or
``dependent'' variables. In observational studies, variables are either
``contributor'' (independent) or ``outcome'' (dependent) variables. If
variables are neither independent nor dependent, they are treated as
co-variates.

\subsection{Assumptions}
Users' assumptions based on domain knowledge are critical for
conducting and contextualizing studies and analyses. Often, users'
assumptions are particular to variables and specific properties (e.g.,
equal variances across different groups). Current tools generally do
not require that users encode these assumptions, leaving them implicit.

Tea takes the opposite approach to contextualize and increase the
transparency of analyses. It requires that users be explicit about
assumptions and statistical properties pertaining to the analysis as a
whole (e.g., acceptable Type I error rate/significance threshold) and
the data.

\subsection{Hypotheses}
\teaHypotheses
Hypotheses drive the statistical analysis process. Users often have
hypotheses that are technically alternative hypotheses. 

Tea focuses on capturing users' alternative hypotheses about the
relationship between two or more variables. Tea uses the alternate
hypothesis to conduct either a two-sided or one-sided statistical
test. By default, Tea uses the null hypothesis that there is no
relationship between variables.
 
Figure~\ref{fig:teaHypotheses} exemplifies the range of hypotheses Tea supports.

\section{Tea's Runtime System} \label{sec:TeaRS}
\tableTeaTests

Tea compiles programs into logical constraints about the data and
variables, which it resolves using a constraint solver. A significant
benefit of using a constraint solver is extensibility. Adding new
statistical tests does not require modifying the core of Tea's runtime
system. Instead, defining a new test requires expressing a single new
logical relationship between a test and its preconditions.

At runtime, Tea invokes a solver that operates on the logical
constraints it computes to produce a list of valid statistical tests
to conduct. This process presents three key technical challenges: (1)
incorporating statistical knowledge as constraints, (2) expressing
user assumptions as constraints, and (3) recursively selecting
statistical tests to verify preconditions of other statistical tests.

% Specifically, Tea uses a
% counter-example guided inductive synthesis~\cite{solar2006combinatorial}
% approach.

\subsection{SMT Solver}
As its constraint solver, Tea uses Z3~\cite{de2008z3}, a Satisfiability Modulo Theory (SMT) solver. 

Satisfiability is the process of finding an assignment to variables that makes a
logical formula true. For example, given the logical rules $0 < x < 100$ and $y
< x$, \{$x = 1, y = 0$\}, \{$x = 10, y = 5$\}, and \{$x = 99, y = -100$\} would all be
valid assignments that satisfy the rules. SMT solvers determine the
satisfiability of logical formulas, which can encode boolean, integer, real
number, and uninterpreted function constraints over variables. SMT solvers can also
be used to encode constraint systems, as we use them here. SMT solvers have been employed in a wide variety of
applications ranging from the synthesis of novel interface
designs~\cite{swearngin2018scout}, the verification of website
accessibility~\cite{panchekha2018verifying}, and the synthesis of data
structures~\cite{loncaric2016cozy}.

\subsection{Logical Encodings}
The first challenge of framing statistical test selection as a constraint satisfaction
problem is defining a logical formulation of statistical
knowledge. 

Tea encodes the applicability of a statistical test based on its preconditions.
A statistical test is applicable if and only if all of its preconditions (which
are properties about variables) hold. We derived preconditions for tests 
from courses~\cite{klemmerCoursera}, statistics
textbooks~\cite{field2012discoveringR}, and publicly available data science
resources from universities~\cite{ucla:whatstat, kent:tutorials}.
 
Tea represents each precondition for a statistical test an uninterpreted
function representing a property over one or more variables. Each property is
assigned \texttt{true} if the property holds for the variable/s; similarly, if the
property does not hold, the property function is assigned \texttt{false}.

% Tea's runtime system uses boolean encodings and propositional formulas to
% synthesize a set of valid statistical tests to test a hypothesis.

Tea also encodes statistical knowledge about variable types and properties that
are essential to statistical analysis as axioms, such as the constraint that only a
continuous variable can be normally distributed. 

\subsection{Algorithm}
Tea frames the problem of finding a set of valid statistical tests as a maximum
satisfiability (MaxSAT) problem that is seeded with user assumptions. 

% Users can provide complete or partial sets of
% assumptions. A complete set of assumptions would include assumptions about all
% possible properties that any test would require. More realistically, users have
% assumptions about only a subset of all properties.  

% Pre-computing all the properties a user does not assume is feasible but time and
% memory intensive especially when the data sets are very large. Instead, Tea uses
% a counter-example guided inductive synthesis~\cite{solar2006combinatorial}
% procedure. This approach separates synthesis of valid tests from verification that the data
% does indeed satisfy a test's preconditions. 

Tea translates each user assumption into an axiom about a property and variable.
For each new statistical test Tea tries to satisfy, Tea verifies if any
precondition of the test violates users' assumptions. If the test's
preconditions do not violate users' assumptions, Tea checks to see if the
precondition holds. For each precondition checked, Tea adds the property and
variable checked as an axiom to observe as future tests are checked. 
The constraint solver then prunes the search space. 

As a result, Tea does not compute all the properties for all variables,
which represents a significant optimization when analyzing
very large datasets.

At the end of this process, Tea finds a set of valid statistical tests
to execute. If this set is empty, Tea defaults to its implementations
of bootstrapping~\cite{efron1992bootstrap}. Otherwise, Tea proceeds
and executes all valid statistical tests. Tea returns a table of
results to users, applying multiple comparison corrections~\cite{holm1979simple} and
calculating effect sizes when appropriate.

\subsection{Optimization: Recursive Queries}
When Tea verifies a property holds for a variable, it often must invoke another
statistical test. For example, to check that two groups have equal variance, 
Tea must execute Levene's test. The statistical test used for
verification may then itself have a precondition, such as a minimum sample size.

Such recursive queries are inefficient for SMT solvers like Z3 to reason
about. To eliminate recursion, Tea lifts some statistical tests to properties.
For instance, Tea does not encode the Levene's test as a statistical test.
Instead, Tea encodes the property of having equal variance between groups and
executes the Levene's test for two groups when verifying that property for particular variables. 

% Another optimization Rooibos makes is to lift some statistical tests to properties. 
% For instance, determining if the variances are equal between groups requires 
% conducting a test such as the Levene's test. To remove recursion from the logical queries, 
% certain tests are encoded as properties instead. Eliminating recurision simplifies  
% the final symbolic query Z3 must satisfy. 
 
% \subsection{Inductive Synthesis in Tea}

% Use reals rather than booleans
% for more complex reasoning, such as for Bayesian inference (??)

% An
% advantage of an SMT solver to if/else statements is that extending Tea by adding
% a statistical test would involve declaring a set of properties that must hold
% about the data before using the test rather than navigating a cascade of nested
% if/else statements that check for the properties. 

% \figureRTeaComparison

\def\r{Pearson's r\xspace}
\def\ktau{Kendall's $\tau$\xspace}
\def\srho{Spearman's $\rho$\xspace}
\def\pb{Pointbiserial\xspace}
\def\student{Student's t-test\xspace}
\def\paired{Paired t-test\xspace}
\def\mannu{Mann-Whitney U\xspace}
\def\wilcox{Wilcoxon signed rank\xspace}
\def\welch{Welch's t-test\xspace}
\def\f{F-test\xspace}
\def\rm{Repeated measures one way ANOVA\xspace}
\def\kw{Kruskal Wallis\xspace}
\def\friedman{Friedman\xspace}
\def\facANOVA{Factorial ANOVA\xspace}
\def\twoANOVA{Two-way ANOVA\xspace}
\def\chiSq{Chi Square\xspace}
\def\fisher{Fisher's Exact\xspace}
\def\boot{Bootstrap\xspace}

\begin{table*}[htbp]
    \begin{center}
    \caption{Results of applying Tea to 12 textbook tutorials.\label{tab:results}}
\begin{minipage}{\linewidth}
\vspace*{-12pt}
\small{%
Tea can prevent false positive and false negative results by suggesting only
tests that satisfy all assumptions.
\textit{Tutorial} gives the test described in the textbook;
\textit{Candidate tests (p-value)} gives all tests a user could run on the provided data with corresponding p-values; 
\textit{Assumptions} gives all satisfied and violated assumptions;
\textit{Tea suggests} indicates which tests Tea suggests based on their assumptions.
Highlighted p-values indicate instances where a candidate test leads to a
wrong conclusion about statistical significance.}
\end{minipage}

    \begin{tabular}{lllc}
    \toprule
    \colH{Tutorial} & \colH{Candidate tests (p-value)} & \colH{Assumptions*} & \colH{Tea suggests} \\
    \midrule
      Pearson                   & \r    \hfill (6.96925e-06) & \valid{2} \valid{4} \invalid{5} & \no \\
      \cite{kabacoff2011action} & \ktau \hfill (2.04198e-05) & \valid{2} \valid{4}             & \yes \\
                                & \srho \hfill (2.83575e-05) & \valid{2} \valid{4}             & \yes \\
    \midrule
    \srho                       & \srho \hfill (.00172)  & \valid{2} \valid{4}           & \yes \\
    \cite{field2012discoveringR} & \r    \hfill (.01115)    &   \valid{2} \invalid{4}                   & \no \\ 
                                & \ktau \hfill (.00126) & \valid{2} \valid{4} & \yes \\
    \midrule
    \ktau                       & \ktau \hfill (.00126) & \valid{2} \valid{4}             & \yes \\
    \cite{field2012discoveringR} & \r   \hfill (.01115) & \valid{2} \invalid{4}             & \no \\
                                & \srho \hfill (.00172) & \valid{2} \valid{4}             & \yes \\
    \midrule
    \pb                         & \pb (\r)  \hfill (.00287) & \valid{2} \valid{4} \invalid{5}             & \no \\
    \cite{field2012discoveringR} & \srho \hfill (.00477) & \valid{2} \invalid{4}             & \no \\
                                 & \ktau \hfill (.00574) & \valid{2} \invalid{4}             & \no \\
                                 &\boot     \hfill (<0.05)                 &           & \yes \\
    \midrule
    \student                     & \student \hfill (.00012) & \valid{2} \valid{4} \valid{5} \valid{6} \valid{7} \valid{8} & \yes \\
    % \cite{kabacoff2011action}    & \paired  \hfill (N/A)                    & \valid{2} & \no \\
    \cite{kabacoff2011action}    & \mannu   \hfill (9.27319e-05)  & \valid{2} \valid{4} \valid{7} \valid{8} & \yes \\
                                %  & \wilcox  \hfill (N/A)                    & \valid{2} & \no \\
                                 & \welch   \hfill (.00065)  & \valid{2} \valid{4} \valid{5} \valid{7} \valid{8} & \yes \\
    \midrule
    \paired                      & \paired \hfill (.03098)    & \valid{2} \valid{4} \valid{5} \valid{7} \valid{8} & \yes \\
    \cite{field2012discoveringR} & \student \hfill (\textbf{.10684})    & \valid{2} \valid{4} \valid{5} \invalid{7} & \no \\
                                 & \mannu   \hfill (\textbf{.06861})    & \valid{2} \valid{4} \invalid{7} & \no \\
                                 & \wilcox  \hfill (.04586)   & \valid{2} \valid{4} \valid{7} \valid{8} & \yes \\
                                 & \welch   \hfill (\textbf{.10724})    & \valid{2} \invalid{7} & \no \\
    \midrule
    \wilcox                      & \wilcox  \hfill (.04657)    & \valid{2} \valid{4} \valid{7} \valid{8} & \yes \\
    \cite{field2012discoveringR} & \student \hfill (.02690)   & \valid{2} \valid{4} \invalid{7} & \no \\
                                 & \paired  \hfill (.01488)   & \valid{2} \valid{4} \invalid{5} \valid{7} \valid{8} & \no \\
                                 & \mannu   \hfill (.00560)   & \valid{2} \valid{4} \invalid{7} & \no \\
                                 & \welch   \hfill (.03572)   & \valid{2} \valid{4} \invalid{7} & \no \\
    \midrule 
    \f                            & \f      \hfill (9.81852e-13)            & \valid{2} \valid{4} \valid{5} \valid{6} \valid{9} & \yes \\
    \cite{field2012discoveringR}  & \kw     \hfill (2.23813e-07)  & \valid{2} \valid{4} \valid{9} & \yes \\ 
                                  & \friedman \hfill (8.66714e-07) & \valid{2} \invalid{7} & \no \\
                                %   & \rm     \hfill ()           & \valid{2}  & \no \\
                                  & \facANOVA \hfill (9.81852e-13)          & \valid{2} \valid{4} \valid{5} \valid{6} \valid{9} & \yes \\
    \midrule
    \kw                           & \kw      \hfill (.03419)    & \valid{2} \valid{4} \valid{9} & \yes \\
    \cite{field2012discoveringR}  & \f       \hfill (\textbf{.05578})              & \valid{2} \valid{4} \invalid{5} \valid{9} & \no \\
                                  & \friedman \hfill (3.02610e-08) & \valid{2} \invalid{7} & \no \\
                                %   & \rm     \hfill ()           & \valid{2} & \no \\
                                  & \facANOVA \hfill (\textbf{.05578})             & \valid{2} \valid{4} \invalid{5} \valid{9} & \no \\
    \midrule
    \rm                           & \rm     \hfill (.0000)           & \valid{2} \valid{4} \valid{5} \valid{6} \valid{7} \valid{9} & \yes \\
    \cite{field2012discoveringR}  & \kw     \hfill (4.51825e-06)  & \valid{2} \valid{4} \invalid{7} \valid{9} & \no \\
                                  & \f      \hfill (1.24278e-07)           & \valid{2} \valid{4} \valid{5} \valid{6} \invalid{7} \valid{9} & \no \\
                                  & \friedman \hfill (5.23589e-11) & \valid{2} \valid{4} \valid{7} \valid{9} & \yes \\ % friedman says 3+ groups
                                  & \facANOVA \hfill (1.24278e-07)          & \valid{2} \valid{4} \valid{5} \valid{6} \valid{9} & \yes \\
    \midrule
    \twoANOVA                     &\twoANOVA \hfill (3.70282e-17)           & \valid{2} \valid{4} \invalid{5} \valid{9} & \no \\
    \cite{field2012discoveringR}  &\boot     \hfill (<0.05)                 &           & \yes \\
    \midrule
    \chiSq                        & \chiSq \hfill (4.76743e-07) & \valid{2} \valid{4} \valid{9} & \yes \\
    \cite{field2012discoveringR}  & \fisher \hfill (4.76743e-07) & \valid{2} \valid{4} \valid{9} & \yes \\
    \bottomrule
    \multicolumn{4}{p{\linewidth}}{*\small{\assume{1} one variable,
                       \assume{2} two variables,
                       \assume{3} two or more variables,
                       \assume{4} continuous vs. categorical vs. ordinal data,
                       \assume{5} normality,
                       \assume{6} equal variance,
                       \assume{7} dependent vs. independent observations,
                       \assume{8} exactly two groups,
                       \assume{9} two or more groups}}

    \end{tabular}
    \end{center}
\end{table*}

\section{Initial Evaluation} \label{sec:eval}
We assessed the benefits of Tea in two ways. First, we compared Tea's
suggestions of statistical tests to suggestions in textbook tutorials.
We use these tutorials as a proxy for expert test selection.
Second, for each tutorial, we compared the analysis results of the test(s)
suggested by Tea to those of the test suggested in the textbook as well as all
other candidate tests. We use the set of all candidate tests as as a proxy for
non-expert test selection.

We differentiate between candidate tests and valid tests. A candidate test can be
computed on the data, when ignoring any preconditions regarding the data types or
distributions. A valid test is a candidate test for which all preconditions are
satisfied.

\subsection{How does Tea compare to textbook tutorials?}
Our goal was to assess how Tea's test selection compared to tests experts
would recommend. 

We sampled 12 data sets and examples from R tutorials (\cite{kabacoff2011action}
and~\cite{field2012discoveringR}). These included eight parametric tests, four
non-parametric tests, and one Chi-square test. We chose these tutorials because they
appeared in two of the top 20 statistical textbooks on Amazon and had publicly available
data sets, which did not require extensive data wrangling.
% and examples that tested the 16 statistical tests that Tea currently supports
%(see~\ref{tab:tea_tests}).

For nine out of the 12 tutorials, Tea suggested the same statistical test
(see Table~\ref{tab:results}). For three out of 12 tutorials, which used a parametric
test, Tea suggested using a non-parametric alternative instead.
The reason for Tea suggesting a non-parametric test was non-normality of the
data. Tea's recommendation of using
a non-parametric test instead of a parametric one did not change the statistical
significance of the result at the $.05$ level.

For the two-way ANOVA tutorial from~\cite{field2012discoveringR}, which studied how gender
and drug usage of individuals affected their perception of attractiveness, a
precondition of the two-way ANOVA is that the dependent measure is normally
distributed in each category. This precondition was violated.  As a result, Tea
defaulted to bootstrapping the means for each group and reported the means and
confidence intervals. 
For the pointbiserial correlation tutorial from~\cite{field2012discoveringR},
Tea also defaulted to bootstrap for two reasons. First, the precondition of
normality is violated. Second, the data uses a dichotomous (nominal) variable,
which renders both \srho and \ktau as invalid.

\begin{comment}

    Limitations and future work??: 
Tea is conservative in its test selection because the \TeaRS will execute a
statistical tests only if all the preconditions are met. Some preconditions may
be more important than others. Future work can consider weighting the
preconditions.
\end{comment}

\subsection{How does Tea compare to non-expert users?}
Our goal was to assess whether any of the tests suggested by Tea (i.e., valid
candidate tests) or any of the invalid candidate tests would lead to a different
conclusion than the one drawn in the tutorial. Table~\ref{tab:results} shows the
results. Specifically, highlighted p-values indicate instances for which the
result of a test differs from the tutorial in terms of statistical significance
at the $.05$ level.

For all of the 12 tutorials, Tea's suggested tests led to the same conclusion
about statistical significance. For two out of the 12 tutorials, two or more
candidate tests led to a different conclusion. These candidate tests were
invalid due to violations of independence or normality.

% captured the complexity (bivariate and multivariate), 
% popularity, 

\section{Limitations and Future Work}
% Could add something about providing support for encoding new tests with less SMT knowledge?
% Tea can return many alternatives. To direct users towards more powerful methods, could calcuclate power and rank for them. 
% Data and analysis provenance through the explicit statement of assumptions

The goal of this paper was to design and assess Tea's high-level DSL
and constraint-based runtime system. Here, we identify limitations of
the current work that suggest opportunities for future work.

\textbf{Empirical evaluation of usability.} While we believe that
abstracting away statistical tests---thus obviating the need for
detailed statistical knowledge---will make Tea substantially easier to
use than conventional statistical tools, an empirical evaluation with
non-statistical expert users will be required to establish this. A
study comparing its use with conventional statistical analysis tools
such as SPSS or R would be of particular interest.

\textbf{Relaxing Tea's conservatism.} Tea is conservative in its test
selection because \TeaRS will execute a statistical test only when all
the preconditions are met. In practice, some preconditions may be more
important than others. For instance, Tea could allow some degree of
deviation from absolute normality. Further evaluation with statistical
and domain experts could help refine Tea's decision making
procedure. %Future work can consider weighting the preconditions.

% Tea is intended to fit into existing data science workflows. To do so, Tea may need to
% accept more file formats than CSVs or provide integration with data
% management systems and databases. \chasins{is this paragraph actually important?}

\textbf{Expanding beyond NHST.}  Tea's architecture is designed to be
flexible and support extension. Currently, Tea provides a module for
Null Hypothesis Significance Testing because NHST is the most common
paradigm in HCI. As statistics norms change, it will be important for
Tea to support a broader range of analyses, including regression and
Bayesian inference.

Extending Tea's architecture and language to Bayesian inference
presents several key research challenges: (1)~easing the process of
choosing and expressing priors, (2)~easing the process of choosing and
expressing models, and (3)~suggesting appropriate statistical tests. A
variety of probabilistic programming languages emphasize language
abstractions that let programmers succinctly express priors and
models--- BUGS~\cite{bugs}, BLOG~\cite{blog}, Stan~\cite{stan}, Church
\cite{church}, and Figaro~\cite{figaro} are a few prominent
examples. Some existing work suggests appropriate statistical tests
for a researcher's
goals~\cite{doingbayesiananalysis,thebayesiannewstatistics,atutorialonpractical},
but these suggestions are generally not embodied in a tool, language,
or programming environment; we look forward to developing ways to
encode these into Tea.

% Extension to other statitical
% methods is important to Tea. 

% \ej{mention something about executing all valid tests could mean that Tea is complicit in multiple testing??}

% It has been a long awaited time for statisitcal analysis to get called out in HCI. 
% The need to improve statistical analysis has received much public attention recently. 
% We offer Tea as one way to improve the state of statistical analysis and interpretation.
% Tea offers an alternative workflow (human and automation) and implementation of tools for future tools. 
% Tea currently supports NHST even though NHST has come under close scrutiny because that is 
% what is common now. The aim of Tea is to offer a new technique and method of improving 
% statisitcal practice. 
% https://www.vox.com/latest-news/2019/3/22/18275913/statistical-significance-p-values-explained

\section{Discussion}
This paper introduces Tea, a high-level programming language that supports users in formalizing and automating statistical analysis.

\textbf{Towards Task-Appropriate Analyses.} Our evaluation shows that Tea's
constraint-based system to find suitable statistical tests generally matches the
choices of experts. In particular, it automatically switches to non-parametric
tests when parametric assumptions are not met. When assumptions are not met, Tea
will always default to tests with fewer assumptions, all the way to the
bootstrap~\cite{efron1992bootstrap}. Tea prevents conducting statistical
analyses that rely on unfounded assumptions. Given Tea's automated test selection
and assumption checking, analyses are more likely to be sound than is currently the
case~\cite{cairns2007hci}.  

% Interestingly, we found that it
% did so in XXX\% of cases.. 

% \kr{more results?}\chasins{explain implications.
% many analyses relying on wrong assumptions for example?}. 

\textbf{Towards Reproducible Analyses.} Researchers have suggested
automation as an opportunity to increase the transparency and
reproducibility of scientific experiments and
findings~\cite{reinhart2015statistics}. Tea programs are relatively
straightforward to write and read and therefore could serve as a way
for researchers to share their analysis for others to reproduce and to
extend. While almost all previous tools place the burden on users to
select suitable statistical tests and check their assumptions, most
users conducting data analysis are not statistical experts.

\textbf{Towards Trustworthy Analyses.} Pre-registration holds the promise of
promoting trustworthy analyses---e.g., by eliminating HARKing, p-hacking, and
cherry picking--- but progress towards mainstream pre-registration has stalled
without a standard format for expressing
study design, hypotheses, and researcher assumptions. Since Tea programs express variables of interest, 
study design, assumptions, and hypotheses, Tea constitutes a potential standard format for pre-registering studies and hypotheses.

\textbf{Fine-Tuning the Division of Labor.} Tea provides what Heer refers to as ``shared representations,'' 
representations that support both human
agency and system automation~\cite{heer2019agency} in statistical analysis.
Users are in ultimate control with Tea. Tea's language empowers users to 
represent their knowledge and intent in conducting analyses (i.e., to test a hypothesis). Users convey
their experimental designs, assumptions, and hypotheses, the high-level goals and domain knowledge
that only the user can provide. Tea takes on the laborious and
error-prone task of searching the space of all possible statistical tests to
evaluate a user-defined hypothesis. Thus, Tea plays a complementary role to users in their efforts to
conduct valid statistical analyses. 

%The declarative nature of Tea could make Tea programs a standard format for pre-registering studies and hypotheses. 
%With Tea, users focus on expressing experimental designs and
%hypotheses to drive the statistical analysis. Users must also make their
%assumptions about the data, variables, and statistical significance levels
%explicit before analyzing the data. Tea could be a tool to support reproducible
%experiments. 

\begin{comment}
%The database community has developed techniques and systems to support  data analysis and management. As a domain-specific programming language, Tea provides an intermediate 
representation that could connect novel interfaces developed in the HCI 
community with advances in databases to create end-to-end systems 
designed to advance the state of statistical analysis~\cite{orr2017probabilistic} 
in fields beyond HCI. For example, Tea provides a programmatic representation 
that could be reused and extended in systems designed 
for executing and exploring multiple analyses for the same experiment 
and hypothesis in parallel, such as interactive explainable 
multiverse analysis reports~\cite{dragicevic2019increasing}.
\end{comment}

\section{Related Work} \label{sec:relatedWork}
Tea extends prior work on domain-specific languages for the data lifecycle, 
tools for statistical analysis, and constraint-based approaches in HCI. 

\subsection{Domain-specific Languages for the Data Lifecycle}
Prior domain-specific languages (DSLs) has focused on several different stages of data exploration, experiment design, and data cleaning to shift the burden of accurate processing from users to systems. To support data exploration, Vega-lite~\cite{satyanarayan2017vega} is a high-level declarative language that supports users in developing interactive data visualizations without writing functional reactive components. PlanOut~\cite{bakshy2014planout} is a DSL for expressing and coordinating online field 
experiments. More niche than PlanOut, Touchstone2 provides the Touchstone Language for specifying 
condition randomization in experiments (e.g., Latin
Squares)~\cite{eiselmayer2019touchstone2}.%Experimental design is also an essential aspect of the domain knowledge users encode in Tea programs. 
To support rapid data cleaning,  Wrangler~\cite{kandel2011wrangler} combines a mixed-initiative interface with a declarative transformation language. Tea can be integrated with tools such as Wrangler that produce cleaned CSV files ready for analysis.

In comparison to these previous DSLs, Tea provides a language to support another crucial step in the data lifecycle: statistical analysis. 

%As a declarative language, Tea has a similar goal for statistical analysis. Tea users do not write any code that performs statistical procedures. They instead focuses on expressing their experimental designs, assumptions, and hypotheses with variables in their data. 

\subsection{Tools for Statistical Analysis}
Research has also introduced tools support statistical analysis in diverse domains. ExperiScope~\cite{guimbretiere2007experiscope} supports users in analyzing complex data logs for interaction techniques. ExperiScope surfaces patterns in the data that would be difficult to detect manually and enables researchers to collect noisier data in the wild that have greater external validity. Touchstone~\cite{mackay2007touchstone} is a comprehensive tool that supports the design and launch of online experiments. Touchstone provides suggestions for data analysis based on experimental design. Touchstone2~\cite{eiselmayer2019touchstone2} builds upon Touchstone and provides
more extensive guidance for evaluating the impact of experimental design on
statistical power. Statsplorer~\cite{wacharamanotham2015statsplorer} is an
educational web application for novices learning about statistics. While more focused on visualizing various alternatives for statistical tests, Statsplorer also automates test selection (for a limited number of statistical tests and by executing simple switch statements) and the checking of assumptions (though it is currently limited to tests of normality and equal variance).~\citet{wacharamanotham2015statsplorer} found that Statsplorer helps HCI students perform better in a subsequent statistics lecture. 

In comparison to Statsplorer, Tea is specifically  designed to integrate into existing workflows (e.g., it can be executed in any Python notebook). It enables reproducing and extending analyses by being script-based, and the analyses are focused on hypotheses that analysts specify. 

%However, Statsplorer is an educational web application that supports a limited number of statistical tests and grounds the analysis procedure in data visualizations. 

%Tea is a Python-based domain specific language designed to integrate into existing workflows that involve programming. Tea also captures users' domain knowledge, and analyses are focused on hypotheses. Tea currently provides a wider range of statistical tests than Statsplorer and can be extended to support emerging statistical practices. 

\subsection{Constraint-based Systems in HCI}
%\chasins{is this subsection really necessary??}
Languages provide semantic structure and meaning that can be reasoned about automatically. For domains with 
well defined goals, constraint solvers can be a   promising technique. Some of the previous constraint-based systems in HCI have been Draco~\cite{moritz2019formalizing} and SetCoLa~\cite{hoffswell2018setcola}, which 
formalize visualization constraints for graphs. Whereas SetCoLa is specifically
focused on graph layout, Draco formalizes visualization best practices as logical constraints to synthesize new visualizations. With additional logical constraints, the knowledge base can grow, supporting the
continual evolution of design recommendations. 

%Tea is more similar to Draco. Tea codifies tests and their preconditions as constraints. Tea aims to provide an architecture that supports the growth of a statistical analysis knowledge base as communities adopt new statistical best practices and methods. 

Another constraint-based system is Scout~\cite{swearngin2018scout}, a mixed-initiative system that supports
interface designers in rapid prototyping. Designers specify high-level 
constraints based on design concepts (e.g., a profile picture should be more
emphasized than the name), and Scout synthesizes novel interfaces. Scout also uses
Z3's theories of booleans and integer linear arithmetic. %Tea currently uses booleans but could leverage integer arithmetic to increase the expressivity of constraints and statistical tests. 

We extend this prior work by providing the first constraint-based system for statistical analysis. 

\section{Conclusion}
Tea is a high-level domain-specific language and runtime system that automates
statistical test selection and execution. Tea achieves these by applying
techniques and ideas from human-computer interaction, programming languages, and
software engineering to statistical analysis. Our hope is that Tea opens up
possibilities for new tools for statistical analysis, helps researchers in
diverse empirical fields, and resolves a century-old question: ``Which test
should I use to test my hypothesis?''

\section{Using Tea}
Tea is an open-source Python package that users can download using
Pip, a Python package manager. Tea can be used in iPython
notebooks. The source code can be accessed at \url{http://tea-lang.org}.

\bibliographystyle{ACM-Reference-Format}
\bibliography{tea-paper}


\begin{thebibliography}{47}

%%% the Web address.

\ifx \showCODEN    \undefined \def \showCODEN     #1{\unskip}     \fi
\ifx \showDOI      \undefined \def \showDOI       #1{#1}\fi
\ifx \showISBNx    \undefined \def \showISBNx     #1{\unskip}     \fi
\ifx \showISBNxiii \undefined \def \showISBNxiii  #1{\unskip}     \fi
\ifx \showISSN     \undefined \def \showISSN      #1{\unskip}     \fi
\ifx \showLCCN     \undefined \def \showLCCN      #1{\unskip}     \fi
\ifx \shownote     \undefined \def \shownote      #1{#1}          \fi
\ifx \showarticletitle \undefined \def \showarticletitle #1{#1}   \fi
\ifx \showURL      \undefined \def \showURL       {\relax}        \fi
% invisible to TeX
\providecommand\bibfield[2]{#2}
\providecommand\bibinfo[2]{#2}
\providecommand\natexlab[1]{#1}
\providecommand\showeprint[2][]{arXiv:#2}

\bibitem[\protect\citeauthoryear{Association}{Association}{1996}]%
        {APATFSI}
\bibfield{author}{\bibinfo{person}{American~Psychological Association}.}
  \bibinfo{year}{1996}\natexlab{}.
\newblock \bibinfo{title}{Task Force on Statistical Inference}.
\newblock
\newblock
\urldef\tempurl%
\url{https://www.apa.org/science/leadership/bsa/statistical/}
\showURL{%
\tempurl}


\bibitem[\protect\citeauthoryear{Bakshy, Eckles, and Bernstein}{Bakshy
  et~al\mbox{.}}{2014}]%
        {bakshy2014planout}
\bibfield{author}{\bibinfo{person}{Eytan Bakshy}, \bibinfo{person}{Dean
  Eckles}, {and} \bibinfo{person}{Michael~S Bernstein}.}
  \bibinfo{year}{2014}\natexlab{}.
\newblock \showarticletitle{Designing and deploying online field experiments}.
  In \bibinfo{booktitle}{\emph{Proceedings of the 23rd international conference
  on World wide web}}. ACM, \bibinfo{pages}{283--292}.
\newblock


\bibitem[\protect\citeauthoryear{Bruin}{Bruin}{2019}]%
        {ucla:whatstat}
\bibfield{author}{\bibinfo{person}{J. Bruin}.} \bibinfo{year}{2019}\natexlab{}.
\newblock \bibinfo{title}{Choosing the Correct Statistical Test in SAS, Stata,
  SPSS and R}.
\newblock
\newblock
\urldef\tempurl%
\url{https://stats.idre.ucla.edu/other/mult-pkg/whatstat/}
\showURL{%
\tempurl}


\bibitem[\protect\citeauthoryear{Cairns}{Cairns}{2007}]%
        {cairns2007hci}
\bibfield{author}{\bibinfo{person}{Paul Cairns}.}
  \bibinfo{year}{2007}\natexlab{}.
\newblock \showarticletitle{HCI... not as it should be: inferential statistics
  in HCI research}. In \bibinfo{booktitle}{\emph{Proceedings of the 21st
  British HCI Group Annual Conference on People and Computers: HCI... but not
  as we know it-Volume 1}}. British Computer Society,
  \bibinfo{pages}{195--201}.
\newblock


\bibitem[\protect\citeauthoryear{Carpenter, Gelman, D.~Hoffman, Lee, Goodrich,
  Betancourt, Brubaker, Guo, Li, and Riddell}{Carpenter et~al\mbox{.}}{2017}]%
        {stan}
\bibfield{author}{\bibinfo{person}{Bob Carpenter}, \bibinfo{person}{Andrew
  Gelman}, \bibinfo{person}{Matthew D.~Hoffman}, \bibinfo{person}{Daniel Lee},
  \bibinfo{person}{Ben Goodrich}, \bibinfo{person}{Michael Betancourt},
  \bibinfo{person}{Marcus Brubaker}, \bibinfo{person}{Jiqiang Guo},
  \bibinfo{person}{Peter Li}, {and} \bibinfo{person}{Allen Riddell}.}
  \bibinfo{year}{2017}\natexlab{}.
\newblock \showarticletitle{Stan : A Probabilistic Programming Language}.
\newblock \bibinfo{journal}{\emph{Journal of Statistical Software}}
  \bibinfo{volume}{76} (\bibinfo{date}{01} \bibinfo{year}{2017}).
\newblock
\urldef\tempurl%
\url{https://doi.org/10.18637/jss.v076.i01}
\showDOI{\tempurl}


\bibitem[\protect\citeauthoryear{De~Moura and Bj{\o}rner}{De~Moura and
  Bj{\o}rner}{2008}]%
        {de2008z3}
\bibfield{author}{\bibinfo{person}{Leonardo De~Moura} {and}
  \bibinfo{person}{Nikolaj Bj{\o}rner}.} \bibinfo{year}{2008}\natexlab{}.
\newblock \showarticletitle{Z3: An efficient SMT solver}. In
  \bibinfo{booktitle}{\emph{International conference on Tools and Algorithms
  for the Construction and Analysis of Systems}}. Springer,
  \bibinfo{pages}{337--340}.
\newblock


\bibitem[\protect\citeauthoryear{Efron}{Efron}{1992}]%
        {efron1992bootstrap}
\bibfield{author}{\bibinfo{person}{Bradley Efron}.}
  \bibinfo{year}{1992}\natexlab{}.
\newblock \showarticletitle{Bootstrap methods: another look at the jackknife}.
\newblock In \bibinfo{booktitle}{\emph{Breakthroughs in statistics}}.
  \bibinfo{publisher}{Springer}, \bibinfo{pages}{569--593}.
\newblock


\bibitem[\protect\citeauthoryear{Ehrlich}{Ehrlich}{1973}]%
        {ehrlich1973participation}
\bibfield{author}{\bibinfo{person}{Isaac Ehrlich}.}
  \bibinfo{year}{1973}\natexlab{}.
\newblock \showarticletitle{Participation in illegitimate activities: A
  theoretical and empirical investigation}.
\newblock \bibinfo{journal}{\emph{Journal of political Economy}}
  \bibinfo{volume}{81}, \bibinfo{number}{3} (\bibinfo{year}{1973}),
  \bibinfo{pages}{521--565}.
\newblock


\bibitem[\protect\citeauthoryear{Eiselmayer, Wacharamanotham, Beaudouin-Lafon,
  and Mackay}{Eiselmayer et~al\mbox{.}}{2019}]%
        {eiselmayer2019touchstone2}
\bibfield{author}{\bibinfo{person}{Alexander Eiselmayer},
  \bibinfo{person}{Chatchavan Wacharamanotham}, \bibinfo{person}{Michel
  Beaudouin-Lafon}, {and} \bibinfo{person}{Wendy Mackay}.}
  \bibinfo{year}{2019}\natexlab{}.
\newblock \showarticletitle{Touchstone2: An Interactive Environment for
  Exploring Trade-offs in HCI Experiment Design}.
\newblock  (\bibinfo{year}{2019}).
\newblock


\bibitem[\protect\citeauthoryear{Field, Miles, and Field}{Field
  et~al\mbox{.}}{2012}]%
        {field2012discoveringR}
\bibfield{author}{\bibinfo{person}{Andy Field}, \bibinfo{person}{Jeremy Miles},
  {and} \bibinfo{person}{Zo{\"e} Field}.} \bibinfo{year}{2012}\natexlab{}.
\newblock \bibinfo{booktitle}{\emph{Discovering statistics using R}}.
\newblock \bibinfo{publisher}{Sage publications}.
\newblock


\bibitem[\protect\citeauthoryear{Fisher}{Fisher}{1937}]%
        {fisher1937design}
\bibfield{author}{\bibinfo{person}{Ronald~Aylmer Fisher}.}
  \bibinfo{year}{1937}\natexlab{}.
\newblock \bibinfo{booktitle}{\emph{The design of experiments}}.
\newblock \bibinfo{publisher}{Oliver And Boyd; Edinburgh; London}.
\newblock


\bibitem[\protect\citeauthoryear{Goodman, Mansinghka, Roy, Bonawitz, and
  Tenenbaum}{Goodman et~al\mbox{.}}{2008}]%
        {church}
\bibfield{author}{\bibinfo{person}{N.~D. Goodman}, \bibinfo{person}{V.~K.
  Mansinghka}, \bibinfo{person}{D.~M. Roy}, \bibinfo{person}{K. Bonawitz},
  {and} \bibinfo{person}{J.~B. Tenenbaum}.} \bibinfo{year}{2008}\natexlab{}.
\newblock \showarticletitle{Church: a language for generative models}.
\newblock \bibinfo{journal}{\emph{Uncertainty in Artificial Intelligence}}.
\newblock


\bibitem[\protect\citeauthoryear{Guimbreti{\`e}re, Dixon, and
  Hinckley}{Guimbreti{\`e}re et~al\mbox{.}}{2007}]%
        {guimbretiere2007experiscope}
\bibfield{author}{\bibinfo{person}{Fran{\c{c}}ois Guimbreti{\`e}re},
  \bibinfo{person}{Morgan Dixon}, {and} \bibinfo{person}{Ken Hinckley}.}
  \bibinfo{year}{2007}\natexlab{}.
\newblock \showarticletitle{ExperiScope: an analysis tool for interaction
  data}. In \bibinfo{booktitle}{\emph{Proceedings of the SIGCHI conference on
  Human factors in computing systems}}. ACM, \bibinfo{pages}{1333--1342}.
\newblock


\bibitem[\protect\citeauthoryear{Heer}{Heer}{2019}]%
        {heer2019agency}
\bibfield{author}{\bibinfo{person}{Jeffrey Heer}.}
  \bibinfo{year}{2019}\natexlab{}.
\newblock \showarticletitle{Agency plus automation: Designing artificial
  intelligence into interactive systems}.
\newblock \bibinfo{journal}{\emph{Proceedings of the National Academy of
  Sciences}} \bibinfo{volume}{116}, \bibinfo{number}{6} (\bibinfo{year}{2019}),
  \bibinfo{pages}{1844--1850}.
\newblock


\bibitem[\protect\citeauthoryear{Hoffswell, Borning, and Heer}{Hoffswell
  et~al\mbox{.}}{2018}]%
        {hoffswell2018setcola}
\bibfield{author}{\bibinfo{person}{Jane Hoffswell}, \bibinfo{person}{Alan
  Borning}, {and} \bibinfo{person}{Jeffrey Heer}.}
  \bibinfo{year}{2018}\natexlab{}.
\newblock \showarticletitle{SetCoLa: High-Level Constraints for Graph Layout}.
  In \bibinfo{booktitle}{\emph{Computer Graphics Forum}},
  Vol.~\bibinfo{volume}{37}. Wiley Online Library, \bibinfo{pages}{537--548}.
\newblock


\bibitem[\protect\citeauthoryear{Holm}{Holm}{1979}]%
        {holm1979simple}
\bibfield{author}{\bibinfo{person}{Sture Holm}.}
  \bibinfo{year}{1979}\natexlab{}.
\newblock \showarticletitle{A simple sequentially rejective multiple test
  procedure}.
\newblock \bibinfo{journal}{\emph{Scandinavian journal of statistics}}
  (\bibinfo{year}{1979}), \bibinfo{pages}{65--70}.
\newblock


\bibitem[\protect\citeauthoryear{Jones, Oliphant, Peterson,
  et~al\mbox{.}}{Jones et~al\mbox{.}}{2019}]%
        {scipy}
\bibfield{author}{\bibinfo{person}{Eric Jones}, \bibinfo{person}{Travis
  Oliphant}, \bibinfo{person}{Pearu Peterson}, {et~al\mbox{.}}}
  \bibinfo{year}{2001--2019}\natexlab{}.
\newblock \bibinfo{title}{{SciPy}: Open source scientific tools for {Python}}.
\newblock
\newblock
\urldef\tempurl%
\url{http://www.scipy.org/}
\showURL{%
\tempurl}


\bibitem[\protect\citeauthoryear{Kabacoff}{Kabacoff}{2011}]%
        {kabacoff2011action}
\bibfield{author}{\bibinfo{person}{Robert~I Kabacoff}.}
  \bibinfo{year}{2011}\natexlab{}.
\newblock \showarticletitle{R: In Action}.
\newblock  (\bibinfo{year}{2011}).
\newblock


\bibitem[\protect\citeauthoryear{Kandel, Paepcke, Hellerstein, and Heer}{Kandel
  et~al\mbox{.}}{2011}]%
        {kandel2011wrangler}
\bibfield{author}{\bibinfo{person}{Sean Kandel}, \bibinfo{person}{Andreas
  Paepcke}, \bibinfo{person}{Joseph Hellerstein}, {and}
  \bibinfo{person}{Jeffrey Heer}.} \bibinfo{year}{2011}\natexlab{}.
\newblock \showarticletitle{Wrangler: Interactive visual specification of data
  transformation scripts}. In \bibinfo{booktitle}{\emph{Proceedings of the
  SIGCHI Conference on Human Factors in Computing Systems}}. ACM,
  \bibinfo{pages}{3363--3372}.
\newblock


\bibitem[\protect\citeauthoryear{Kaptein and Robertson}{Kaptein and
  Robertson}{2012}]%
        {kaptein2012rethinking}
\bibfield{author}{\bibinfo{person}{Maurits Kaptein} {and} \bibinfo{person}{Judy
  Robertson}.} \bibinfo{year}{2012}\natexlab{}.
\newblock \showarticletitle{Rethinking statistical analysis methods for CHI}.
  In \bibinfo{booktitle}{\emph{Proceedings of the SIGCHI Conference on Human
  Factors in Computing Systems}}. ACM, \bibinfo{pages}{1105--1114}.
\newblock


\bibitem[\protect\citeauthoryear{Kay, Nelson, and Hekler}{Kay
  et~al\mbox{.}}{2016}]%
        {kay2016researcher}
\bibfield{author}{\bibinfo{person}{Matthew Kay}, \bibinfo{person}{Gregory~L
  Nelson}, {and} \bibinfo{person}{Eric~B Hekler}.}
  \bibinfo{year}{2016}\natexlab{}.
\newblock \showarticletitle{Researcher-centered design of statistics: Why
  Bayesian statistics better fit the culture and incentives of HCI}. In
  \bibinfo{booktitle}{\emph{Proceedings of the 2016 CHI Conference on Human
  Factors in Computing Systems}}. ACM, \bibinfo{pages}{4521--4532}.
\newblock


\bibitem[\protect\citeauthoryear{Klemmer and Wobbrock}{Klemmer and
  Wobbrock}{2019}]%
        {klemmerCoursera}
\bibfield{author}{\bibinfo{person}{Scott Klemmer} {and} \bibinfo{person}{Jacob
  Wobbrock}.} \bibinfo{year}{2019}\natexlab{}.
\newblock \bibinfo{title}{Designing, Running, and Analyzing Experiments}.
\newblock
\newblock
\urldef\tempurl%
\url{https://www.coursera.org/learn/designexperiments}
\showURL{%
\tempurl}


\bibitem[\protect\citeauthoryear{Kruschke}{Kruschke}{2010}]%
        {doingbayesiananalysis}
\bibfield{author}{\bibinfo{person}{John~K. Kruschke}.}
  \bibinfo{year}{2010}\natexlab{}.
\newblock \bibinfo{booktitle}{\emph{Doing Bayesian Data Analysis: A Tutorial
  with R and BUGS} (\bibinfo{edition}{1st} ed.)}.
\newblock \bibinfo{publisher}{Academic Press, Inc.}, \bibinfo{address}{Orlando,
  FL, USA}.
\newblock
\showISBNx{0123814855, 9780123814852}


\bibitem[\protect\citeauthoryear{Kruschke and Liddell}{Kruschke and
  Liddell}{2018}]%
        {thebayesiannewstatistics}
\bibfield{author}{\bibinfo{person}{John~K. Kruschke} {and}
  \bibinfo{person}{Torrin~M. Liddell}.} \bibinfo{year}{2018}\natexlab{}.
\newblock \showarticletitle{The Bayesian New Statistics: Hypothesis testing,
  estimation, meta-analysis, and power analysis from a Bayesian perspective}.
\newblock \bibinfo{journal}{\emph{Psychonomic Bulletin {\&} Review}}
  \bibinfo{volume}{25}, \bibinfo{number}{1} (\bibinfo{date}{01 Feb}
  \bibinfo{year}{2018}), \bibinfo{pages}{178--206}.
\newblock
\showISSN{1531-5320}
\urldef\tempurl%
\url{https://doi.org/10.3758/s13423-016-1221-4}
\showDOI{\tempurl}


\bibitem[\protect\citeauthoryear{Libraries}{Libraries}{2019}]%
        {kent:tutorials}
\bibfield{author}{\bibinfo{person}{Kent State~University Libraries}.}
  \bibinfo{year}{2019}\natexlab{}.
\newblock \bibinfo{title}{SPSS Tutorials: Analyzing Data}.
\newblock
\newblock
\urldef\tempurl%
\url{https://libguides.library.kent.edu/SPSS/AnalyzeData}
\showURL{%
\tempurl}


\bibitem[\protect\citeauthoryear{Loncaric, Torlak, and Ernst}{Loncaric
  et~al\mbox{.}}{2016}]%
        {loncaric2016cozy}
\bibfield{author}{\bibinfo{person}{Calvin Loncaric}, \bibinfo{person}{Emina
  Torlak}, {and} \bibinfo{person}{Michael~D Ernst}.}
  \bibinfo{year}{2016}\natexlab{}.
\newblock \showarticletitle{Fast synthesis of fast collections}.
\newblock \bibinfo{journal}{\emph{ACM SIGPLAN Notices}} \bibinfo{volume}{51},
  \bibinfo{number}{6} (\bibinfo{year}{2016}), \bibinfo{pages}{355--368}.
\newblock


\bibitem[\protect\citeauthoryear{Lunn, Thomas, Best, and Spiegelhalter}{Lunn
  et~al\mbox{.}}{2000}]%
        {bugs}
\bibfield{author}{\bibinfo{person}{David~J. Lunn}, \bibinfo{person}{Andrew
  Thomas}, \bibinfo{person}{Nicky Best}, {and} \bibinfo{person}{David
  Spiegelhalter}.} \bibinfo{year}{2000}\natexlab{}.
\newblock \showarticletitle{WinBUGS - A Bayesian modelling framework: Concepts,
  structure, and extensibility}.
\newblock \bibinfo{journal}{\emph{Statistics and Computing}}
  \bibinfo{volume}{10}, \bibinfo{number}{4} (\bibinfo{date}{01 Oct}
  \bibinfo{year}{2000}), \bibinfo{pages}{325--337}.
\newblock
\showISSN{1573-1375}
\urldef\tempurl%
\url{https://doi.org/10.1023/A:1008929526011}
\showDOI{\tempurl}


\bibitem[\protect\citeauthoryear{Mackay, Appert, Beaudouin-Lafon, Chapuis, Du,
  Fekete, and Guiard}{Mackay et~al\mbox{.}}{2007}]%
        {mackay2007touchstone}
\bibfield{author}{\bibinfo{person}{Wendy~E Mackay}, \bibinfo{person}{Caroline
  Appert}, \bibinfo{person}{Michel Beaudouin-Lafon}, \bibinfo{person}{Olivier
  Chapuis}, \bibinfo{person}{Yangzhou Du}, \bibinfo{person}{Jean-Daniel
  Fekete}, {and} \bibinfo{person}{Yves Guiard}.}
  \bibinfo{year}{2007}\natexlab{}.
\newblock \showarticletitle{Touchstone: exploratory design of experiments}. In
  \bibinfo{booktitle}{\emph{Proceedings of the SIGCHI conference on Human
  factors in computing systems}}. ACM, \bibinfo{pages}{1425--1434}.
\newblock


\bibitem[\protect\citeauthoryear{Masson}{Masson}{2011}]%
        {atutorialonpractical}
\bibfield{author}{\bibinfo{person}{Michael E.~J. Masson}.}
  \bibinfo{year}{2011}\natexlab{}.
\newblock \showarticletitle{A tutorial on a practical {Bayesian} alternative to
  null-hypothesis significance testing}.
\newblock \bibinfo{journal}{\emph{Behavior Research Methods}}
  \bibinfo{volume}{43}, \bibinfo{number}{3} (\bibinfo{date}{Sept.}
  \bibinfo{year}{2011}), \bibinfo{pages}{679--690}.
\newblock
\showISSN{1554-3528}
\urldef\tempurl%
\url{https://doi.org/10.3758/s13428-010-0049-5}
\showDOI{\tempurl}


\bibitem[\protect\citeauthoryear{Milch, Marthi, Russell, Sontag, Ong, and
  Kolobov}{Milch et~al\mbox{.}}{2005}]%
        {blog}
\bibfield{author}{\bibinfo{person}{Brian Milch}, \bibinfo{person}{Bhaskara
  Marthi}, \bibinfo{person}{Stuart Russell}, \bibinfo{person}{David Sontag},
  \bibinfo{person}{Daniel~L. Ong}, {and} \bibinfo{person}{Andrey Kolobov}.}
  \bibinfo{year}{2005}\natexlab{}.
\newblock \showarticletitle{{BLOG}: Probabilistic Models with Unknown Objects}.
  In \bibinfo{booktitle}{\emph{Proc. 19th International Joint Conference on
  Artificial Intelligence}}. \bibinfo{pages}{1352--1359}.
\newblock
\urldef\tempurl%
\url{http://sites.google.com/site/bmilch/papers/blog-ijcai05.pdf}
\showURL{%
\tempurl}


\bibitem[\protect\citeauthoryear{Moritz, Wang, Nelson, Lin, Smith, Howe, and
  Heer}{Moritz et~al\mbox{.}}{2019}]%
        {moritz2019formalizing}
\bibfield{author}{\bibinfo{person}{Dominik Moritz}, \bibinfo{person}{Chenglong
  Wang}, \bibinfo{person}{Greg~L Nelson}, \bibinfo{person}{Halden Lin},
  \bibinfo{person}{Adam~M Smith}, \bibinfo{person}{Bill Howe}, {and}
  \bibinfo{person}{Jeffrey Heer}.} \bibinfo{year}{2019}\natexlab{}.
\newblock \showarticletitle{Formalizing visualization design knowledge as
  constraints: Actionable and extensible models in Draco}.
\newblock \bibinfo{journal}{\emph{IEEE transactions on visualization and
  computer graphics}} \bibinfo{volume}{25}, \bibinfo{number}{1}
  (\bibinfo{year}{2019}), \bibinfo{pages}{438--448}.
\newblock


\bibitem[\protect\citeauthoryear{Oliphant}{Oliphant}{2006}]%
        {oliphant2006numpy}
\bibfield{author}{\bibinfo{person}{Travis~E Oliphant}.}
  \bibinfo{year}{2006}\natexlab{}.
\newblock \bibinfo{booktitle}{\emph{A guide to NumPy}}.
  Vol.~\bibinfo{volume}{1}.
\newblock \bibinfo{publisher}{Trelgol Publishing USA}.
\newblock


\bibitem[\protect\citeauthoryear{Panchekha, Geller, Ernst, Tatlock, and
  Kamil}{Panchekha et~al\mbox{.}}{2018}]%
        {panchekha2018verifying}
\bibfield{author}{\bibinfo{person}{Pavel Panchekha}, \bibinfo{person}{Adam~T
  Geller}, \bibinfo{person}{Michael~D Ernst}, \bibinfo{person}{Zachary
  Tatlock}, {and} \bibinfo{person}{Shoaib Kamil}.}
  \bibinfo{year}{2018}\natexlab{}.
\newblock \showarticletitle{Verifying that web pages have accessible layout}.
  In \bibinfo{booktitle}{\emph{Proceedings of the 39th ACM SIGPLAN Conference
  on Programming Language Design and Implementation}}. ACM,
  \bibinfo{pages}{1--14}.
\newblock


\bibitem[\protect\citeauthoryear{Pfeffer}{Pfeffer}{2011}]%
        {figaro}
\bibfield{author}{\bibinfo{person}{Avi Pfeffer}.}
  \bibinfo{year}{2011}\natexlab{}.
\newblock \showarticletitle{Practical Probabilistic Programming}. In
  \bibinfo{booktitle}{\emph{Inductive Logic Programming}},
  \bibfield{editor}{\bibinfo{person}{Paolo Frasconi} {and}
  \bibinfo{person}{Francesca~A. Lisi}} (Eds.). \bibinfo{publisher}{Springer
  Berlin Heidelberg}, \bibinfo{address}{Berlin, Heidelberg},
  \bibinfo{pages}{2--3}.
\newblock
\showISBNx{978-3-642-21295-6}


\bibitem[\protect\citeauthoryear{Reinhart}{Reinhart}{2015}]%
        {reinhart2015statistics}
\bibfield{author}{\bibinfo{person}{Alex Reinhart}.}
  \bibinfo{year}{2015}\natexlab{}.
\newblock \bibinfo{booktitle}{\emph{Statistics done wrong: The woefully
  complete guide}}.
\newblock \bibinfo{publisher}{No starch press}.
\newblock


\bibitem[\protect\citeauthoryear{Satyanarayan, Moritz, Wongsuphasawat, and
  Heer}{Satyanarayan et~al\mbox{.}}{2017}]%
        {satyanarayan2017vega}
\bibfield{author}{\bibinfo{person}{Arvind Satyanarayan},
  \bibinfo{person}{Dominik Moritz}, \bibinfo{person}{Kanit Wongsuphasawat},
  {and} \bibinfo{person}{Jeffrey Heer}.} \bibinfo{year}{2017}\natexlab{}.
\newblock \showarticletitle{Vega-lite: A grammar of interactive graphics}.
\newblock \bibinfo{journal}{\emph{IEEE transactions on visualization and
  computer graphics}} \bibinfo{volume}{23}, \bibinfo{number}{1}
  (\bibinfo{year}{2017}), \bibinfo{pages}{341--350}.
\newblock


\bibitem[\protect\citeauthoryear{Seabold and Perktold}{Seabold and
  Perktold}{2010}]%
        {seabold2010statsmodels}
\bibfield{author}{\bibinfo{person}{Skipper Seabold} {and}
  \bibinfo{person}{Josef Perktold}.} \bibinfo{year}{2010}\natexlab{}.
\newblock \showarticletitle{Statsmodels: Econometric and statistical modeling
  with python}. In \bibinfo{booktitle}{\emph{Proceedings of the 9th Python in
  Science Conference}}, Vol.~\bibinfo{volume}{57}. Scipy, \bibinfo{pages}{61}.
\newblock


\bibitem[\protect\citeauthoryear{Swearngin, Ko, and Fogarty}{Swearngin
  et~al\mbox{.}}{2018}]%
        {swearngin2018scout}
\bibfield{author}{\bibinfo{person}{Amanda Swearngin}, \bibinfo{person}{Andrew~J
  Ko}, {and} \bibinfo{person}{James Fogarty}.} \bibinfo{year}{2018}\natexlab{}.
\newblock \showarticletitle{Scout: Mixed-Initiative Exploration of Design
  Variations through High-Level Design Constraints}. In
  \bibinfo{booktitle}{\emph{The 31st Annual ACM Symposium on User Interface
  Software and Technology Adjunct Proceedings}}. ACM,
  \bibinfo{pages}{134--136}.
\newblock


\bibitem[\protect\citeauthoryear{Vandaele}{Vandaele}{1987}]%
        {vandaele1987participation}
\bibfield{author}{\bibinfo{person}{Walter Vandaele}.}
  \bibinfo{year}{1987}\natexlab{}.
\newblock \bibinfo{booktitle}{\emph{Participation in illegitimate activities:
  Ehrlich revisited, 1960}}. Vol.~\bibinfo{volume}{8677}.
\newblock \bibinfo{publisher}{Inter-university Consortium for Political and
  Social Research}.
\newblock


\bibitem[\protect\citeauthoryear{Venables and Ripley}{Venables and
  Ripley}{2013}]%
        {venables2013modern}
\bibfield{author}{\bibinfo{person}{William~N Venables} {and}
  \bibinfo{person}{Brian~D Ripley}.} \bibinfo{year}{2013}\natexlab{}.
\newblock \bibinfo{booktitle}{\emph{Modern applied statistics with S-PLUS}}.
\newblock \bibinfo{publisher}{Springer Science \& Business Media}.
\newblock


\bibitem[\protect\citeauthoryear{Wacharamanotham, Subramanian, Volkel, and
  Borchers}{Wacharamanotham et~al\mbox{.}}{2015}]%
        {wacharamanotham2015statsplorer}
\bibfield{author}{\bibinfo{person}{Chat Wacharamanotham},
  \bibinfo{person}{Krishna Subramanian}, \bibinfo{person}{Sarah~Theres Volkel},
  {and} \bibinfo{person}{Jan Borchers}.} \bibinfo{year}{2015}\natexlab{}.
\newblock \showarticletitle{Statsplorer: Guiding novices in statistical
  analysis}. In \bibinfo{booktitle}{\emph{Proceedings of the 33rd Annual ACM
  Conference on Human Factors in Computing Systems}}. ACM,
  \bibinfo{pages}{2693--2702}.
\newblock


\bibitem[\protect\citeauthoryear{Wickham et~al\mbox{.}}{Wickham
  et~al\mbox{.}}{2014}]%
        {wickham2014tidy}
\bibfield{author}{\bibinfo{person}{Hadley Wickham} {et~al\mbox{.}}}
  \bibinfo{year}{2014}\natexlab{}.
\newblock \showarticletitle{Tidy data}.
\newblock \bibinfo{journal}{\emph{Journal of Statistical Software}}
  \bibinfo{volume}{59}, \bibinfo{number}{10} (\bibinfo{year}{2014}),
  \bibinfo{pages}{1--23}.
\newblock


\bibitem[\protect\citeauthoryear{{Wikipedia contributors}}{{Wikipedia
  contributors}}{2019a}]%
        {wiki:jmp}
\bibfield{author}{\bibinfo{person}{{Wikipedia contributors}}.}
  \bibinfo{year}{2019}\natexlab{a}.
\newblock \bibinfo{title}{JMP (statistical software) --- {Wikipedia}{,} The
  Free Encyclopedia}.
\newblock
  \bibinfo{howpublished}{\url{https://en.wikipedia.org/w/index.php?title=JMP_(statistical_software)&oldid=887217350}}.
\newblock
\newblock
\shownote{[Online; accessed 5-April-2019].}


\bibitem[\protect\citeauthoryear{{Wikipedia contributors}}{{Wikipedia
  contributors}}{2019b}]%
        {wiki:r-language}
\bibfield{author}{\bibinfo{person}{{Wikipedia contributors}}.}
  \bibinfo{year}{2019}\natexlab{b}.
\newblock \bibinfo{title}{R (programming language) --- {Wikipedia}{,} The Free
  Encyclopedia}.
\newblock
  \bibinfo{howpublished}{\url{https://en.wikipedia.org/w/index.php?title=R_(programming_language)&oldid=890657071}}.
\newblock
\newblock
\shownote{[Online; accessed 5-April-2019].}


\bibitem[\protect\citeauthoryear{{Wikipedia contributors}}{{Wikipedia
  contributors}}{2019c}]%
        {wiki:sas}
\bibfield{author}{\bibinfo{person}{{Wikipedia contributors}}.}
  \bibinfo{year}{2019}\natexlab{c}.
\newblock \bibinfo{title}{SAS (software) --- {Wikipedia}{,} The Free
  Encyclopedia}.
\newblock
  \bibinfo{howpublished}{\url{https://en.wikipedia.org/w/index.php?title=SAS_(software)&oldid=890451452}}.
\newblock
\newblock
\shownote{[Online; accessed 5-April-2019].}


\bibitem[\protect\citeauthoryear{{Wikipedia contributors}}{{Wikipedia
  contributors}}{2019d}]%
        {wiki:spss}
\bibfield{author}{\bibinfo{person}{{Wikipedia contributors}}.}
  \bibinfo{year}{2019}\natexlab{d}.
\newblock \bibinfo{title}{SPSS --- {Wikipedia}{,} The Free Encyclopedia}.
\newblock
  \bibinfo{howpublished}{\url{https://en.wikipedia.org/w/index.php?title=SPSS&oldid=888470477}}.
\newblock
\newblock
\shownote{[Online; accessed 5-April-2019].}


\bibitem[\protect\citeauthoryear{Wilkinson}{Wilkinson}{1999}]%
        {wilkinson1999statistical}
\bibfield{author}{\bibinfo{person}{Leland Wilkinson}.}
  \bibinfo{year}{1999}\natexlab{}.
\newblock \showarticletitle{Statistical methods in psychology journals:
  Guidelines and explanations.}
\newblock \bibinfo{journal}{\emph{American psychologist}} \bibinfo{volume}{54},
  \bibinfo{number}{8} (\bibinfo{year}{1999}), \bibinfo{pages}{594}.
\newblock


\end{thebibliography}

\end{document}